\pdfoutput=1
\documentclass[structabstract]{aa}  
\usepackage{graphicx}
\usepackage{txfonts}
\usepackage{longtable,lscape}
\begin{document}
   \title{A proper motion study of the Lupus clouds using VO tools}

   \subtitle{}

 \author{Bel\'en L\'opez Mart\'{\i}
          \inst{1}
          \and
          Francisco Jim\'enez-Esteban \inst{1,2,3}
	  \and 
	  Enrique Solano \inst{1,2}
          }

   \institute{
        Centro de Astrobiolog\'{\i}a (INTA-CSIC), Departamento de Astrof\'{\i}sica, P.O. Box 78, E-28261 Villanueva de la Ca\~nada, Madrid, Spain\\
	\email{belen@cab.inta-csic.es}
	\and
	Spanish Virtual Observatory. Spain
	\and
	Saint Louis University, Madrid Campus, Division of Science and Engineering, Avenida~del~Valle 34, E-28003 Madrid, Spain
	}

   \date{Received 21 December 2010; Accepted 24 February 2011}

 
  \abstract
   {The Lupus dark cloud complex is a well-known, nearby low-mass star-forming region, probably associated with the Gould Belt. In recent years, the number of stellar and substellar Lupus candidate members has been remarkably increased thanks to the \emph{Cores to Disks (c2d) Spitzer} Legacy Program and other studies. However, most of these newly discovered objects still lack confirmation that they belong to the dark clouds.}
   {By using available kinematical information, we test the membership of the new Lupus candidate members proposed by the \emph{c2d} program and by a complementary optical survey. We also investigate the relationship between the proper motions and other properties of the objects, in order to get some clues about their formation and early evolution.}
   {We compiled a list of members and possible members of Lupus~1, 3, and 4, together with all available information on their spectral types, disks, and physical parameters. Using Virtual Observatory tools, we cross-matched this list with the available astrometric catalogues to get proper motions for our objects. Our final sample contains sources with magnitudes $I<16$~mag and estimated masses $\gtrsim0.1M_{\odot}$.}
   {According to the kinematic information, our sources can be divided into two main groups. The first one contains sources with higher proper motions in agreement with other Gould Belt populations and with spatial distribution, optical and near-infrared colours, and disk composition consistent with these objects belonging to the Lupus clouds. In the second group, sources have lower proper motions with random orientations, and they are mostly located outside the cloud cores, making their association with the Lupus complex more doubtful. We investigate the properties of the higher proper motion group, but cannot find any correlations with spatial location, binarity, the presence of a circumstellar disk, or with physical properties such as effective temperature, luminosity, mass, or age.
   }
   {We conclude that the lower proper motion group probably represents a background population or mixture of populations unrelated to the Lupus clouds. The higher proper motion group, on the other hand, has properties consistent with it being a genuine population of the Lupus star-forming region. More accurate proper motions and/or radial velocity information are required for a more detailed study of the kinematic properties of the Lupus stellar members.}


   \maketitle

%
\section{Introduction}\label{sec:intro}

The Lupus dark cloud complex consists of six molecular clouds named Lupus 1 to 6 (Tachihara et al. \cite{tachihara96}, \cite{tachihara01}). In terms of angular extent, it is one of the largest low-mass star-forming complexes in the sky (nearly 20$^{\circ}$ across, with galactic longitudes $334^{\circ}<l<352^{\circ}$ and latitudes $5^{\circ}<b<25^{\circ}$), and it contains one of the richest associations of T~Tauri stars (TTS). The clouds are projected against the Scorpius-Centaurus OB association, a vast complex at an average distance of about 140~pc from the Sun. This is one of the main structures of the Gould Belt, a young, expanding ring-like arrangement of  OB  associations, molecular clouds and other tracers of star formation  that dominates the solar neighbourhood up to a distance of 600 pc  (e.g. Comer\'on et al. \cite{comeron94}; P\"oppel \cite{poppel97}). The estimated distance of 140-220~pc to the Lupus clouds matches that of the Scorpius-Centaurus association, strongly suggesting a physical relationship between both entities. 

The first studies of the stellar content identified at least 65 TTS within the complex, most of them classical T~Tauri stars (CTTS), that is, young stars with indications of ongoing accretion in their spectra (e.g. Schwartz \cite{schwartz77}; Krautter \cite{krautter91}; Hughes et al. \cite{hughes94}). Later studies, the majority of them focused on the Lupus~3 cloud, used different techniques to increase the stellar and substellar census of the complex (e.g. Nakajima et al. \cite{nakajima00}; Comer\'on et al. \cite{comeron03}; L\'opez Mart\'{\i} et al. \cite{lm05}; Allen et al. \cite{allen07}). A peculiarity of the Lupus region is that the distribution of stellar spectral types is dominated by M-type objects, in contrast to other low-mass star-forming regions (e.g. Appenzeller et al. \cite{appenzeller83}; Hughes et al. \cite{hughes94}; Wichmann et al. \cite{wichmann97a}). The only exceptions are the two intermediate-mass Herbig Ae/Be stars HR~5999 and HR~6000 in the Lupus~3 cloud. 

Three of the clouds, namely Lupus~1, 3, and 4, have been the target of  {\em Spitzer} observations within the \emph{cores to disks (c2d)} Legacy Program (Allers et al. \cite{allers06}; Chapman et al. \cite{chapman07}; Mer\'{\i}n et al. \cite{merin08}). This study, one of the most comprehensive surveys in the complex, has confirmed the youth of many previously suspected members of the region. It has also provided a list of about hundred low-mass candidate members based on the detection of excess in their spectral energy distributions (SEDs) at infrared wavelengths, where warm dust dominates the emission. Therefore, most of the objects presented by the {\em c2d} program are surrounded by circum(sub)stellar disks and/or envelopes (so-called ``class I and II sources'').

In a recent paper, Comer\'on, Spezzi \& L\'opez Mart\'{\i} (\cite{comeron09}, hereafter Comer\'on et al. \cite{comeron09}) have reported on a broad-band optical ($RIz$) survey of the Lupus clouds with the WFI mosaic camera at the ESO/MPG 2.2m telescope on La Silla, complemented with $JHKs$ photometry from 2MASS. In their surveyed areas, which overlap almost completely with the areas observed by the  {\em c2d} program, they identify about 130 new low-mass candidate members of Lupus. Although their estimated ages are in the same range as the known members, these new candidate members are only moderately concentrated towards the dark clouds, and very few of the objects show infrared excess (that is, they are mostly so-called ``class III sources''). Thus, the discovery of this new population poses interesting questions about the origin and early evolution of the stellar content of the Lupus complex as a whole. Comer\'on et al. analysed several possible formation scenarios that may have led to the observed populations. They also widely discussed the probability that the new diskless candidate members belong to an unrelated foreground population associated with the Gould Belt, which they held as unlikely given their apparent association with the dark cloud cores.

Kinematic information can help to shed more light on this issue, by providing further evidence of the membership of these objects in the same clustered structure as the known Lupus members. Moreover, the kinematic properties of the members of a given region or association hold important clues about its history. Some formation models predict that the early dynamical evolution of the parent proto-stellar cluster should lead to mass-dependent kinematic distributions (Kroupa \& Bouvier \cite{kroupa03}), while other numerical simulations predict similar kinematic properties over the whole mass spectrum (Bate et al. \cite{bate03}; Bate \cite{bate09}). A number of surveys have used proper motions to  identify and confirm new low-mass members in young associations and clusters (e.g. Moraux et al. \cite{moraux01}; Kraus \& Hillenbrand \cite{kraus07}; Bouy \& Mart\'{\i}n \cite{bouy09}; Caballero \cite{caballero10}), and various authors have used radial velocity measurements to study the kinematic properties of young low-mass objects (e.g. Joergens \& Guenther \cite{joergens01}; Joergens \cite{joergens06}; Jeffries et al. \cite{jeffries06}; Maxted et al. \cite{maxted08}).

In this paper, we make use of proper motions from astrometric catalogues and of Virtual Observatory\footnote{\tt http://www.ivoa.net} (VO) tools to investigate the kinematic properties of the Lupus members and candidates. Our goals are to provide further evidence of a common origin for all these objects, to test their association with the Lupus dark clouds, and to look for correlations between the proper motions and other physical properties of the objects, aiming at a better understanding of the formation of the stellar and substellar population within the Lupus complex.

%
\section{Proper motions of Lupus members and candidate members}\label{sec:pms}

\begin{table}
\caption{Number of objects in each catalogue}             
\label{tab:nobj}   
\centering           
\begin{tabular}{l r r r r}
\hline
\hline
  \multicolumn{1}{l}{Catalogue} &
  \multicolumn{1}{c}{Lupus~1} &
  \multicolumn{1}{c}{Lupus~3} &
  \multicolumn{1}{c}{Lupus~4} &
  \multicolumn{1}{c}{All} \\
\hline  
  Previous$^{(a)}$   & 13 &   43 &    9$^{(b)}$ &   65 \\
  $c2d$                      & 17 & 124 & 18               & 159 \\
  WFI                          & 65 &   72 &    6               & 143 \\
  Compiled$^{(c)}$ & 90 & 208 & 33               & 331 \\
  \hline
  USNO-B1 & 30 &   55 &   8 &   93 \\
  SSS           & 80 & 163 & 23 & 266 \\
  PPMX        & 28 &   21 &   2 &   51 \\
  UCAC3     & 69 &   96 & 13 & 178 \\
\hline\end{tabular}
\begin{flushleft}
{\bf Notes.}\\
$^{(a)}$ Objects from the review by Comer\'on (\cite{comeron08}). \\ 
$^{(b)}$ Comer\'on (\cite{comeron08}) erroneously lists as Lupus~4 members six objects actually belonging to the Norma~1 cloud, which are not considered in the present work. \\
$^{(c)}$ Objects included in several catalogues have been counted once.\\
\end{flushleft}
\end{table}

   \begin{figure*}[t]
   \centering
  \includegraphics[width=16cm]{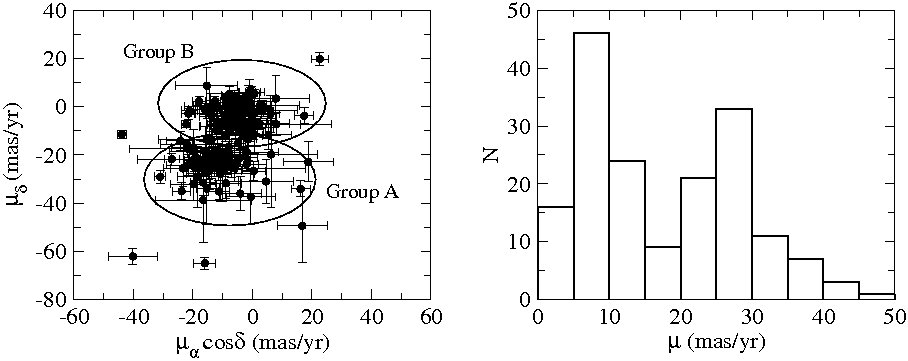}\hfill\\
      \caption{\footnotesize
	       Vector point diagram (left panel) and proper motion histogram (right panel) for the Lupus sources included in the UCAC3 catalogue. A few sources with very large proper motions have not been plotted  (see Table~\ref{tab:outl}).}
         \label{fig:pm}
   \end{figure*}

   \begin{figure*}[ht]
   \centering
    \includegraphics[width=15.7cm]{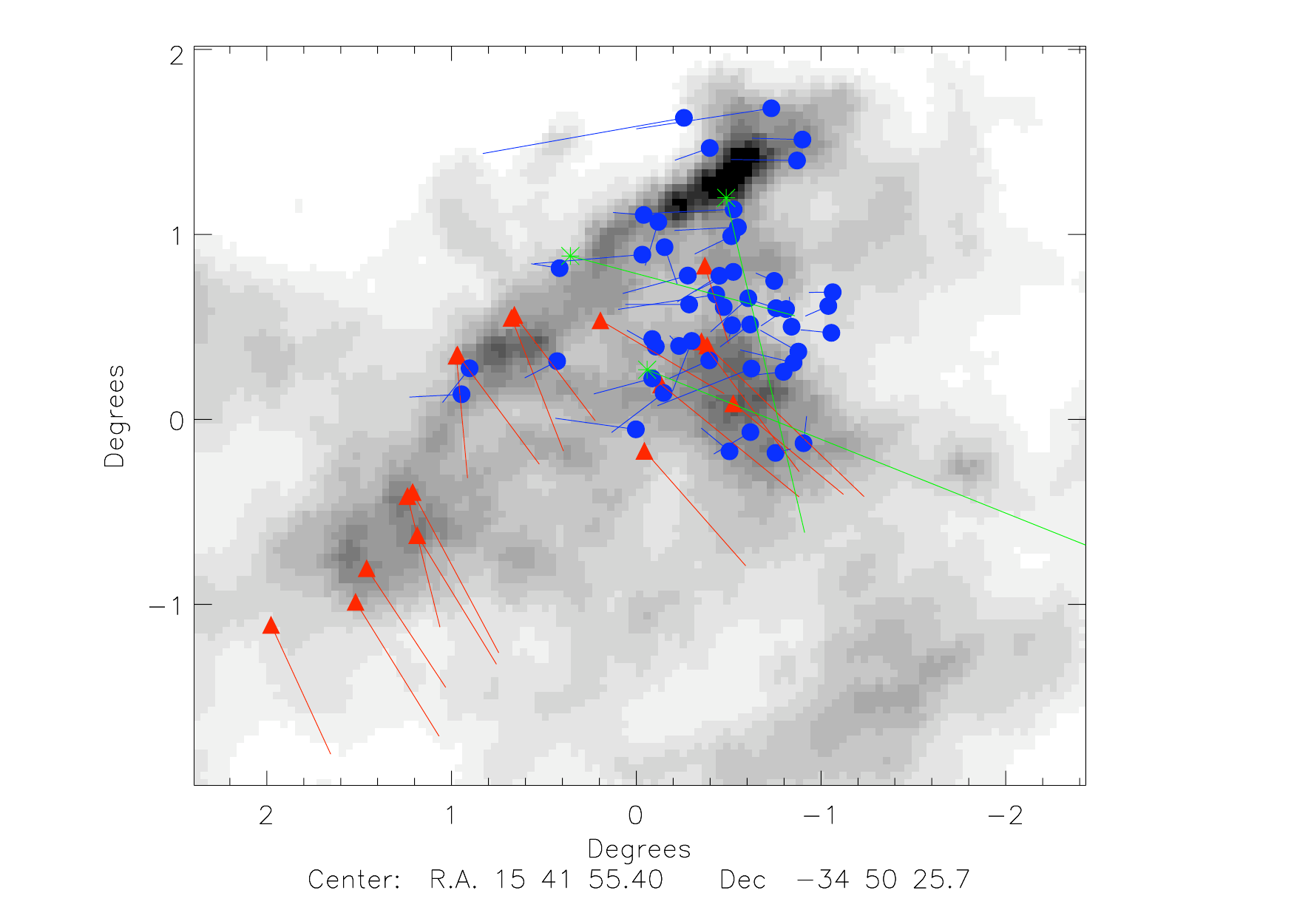}\hfill
       \caption{\footnotesize
	       Current spatial location of the Lupus sources from our UCAC3 sample in Lupus~1. Group A sources are indicated with red triangles, Group B sources with blue circles and outliers with green asterisks. The lines show the expected displacement of these objects within 10$^5$~Myr. The background image is a dust map by Schlegel et al. (\cite{schlegel98}) showing the location of the cloud cores.
	       }
         \label{fig:lup1dist}
   \end{figure*}

   \begin{figure*}[ht]
   \centering
    \includegraphics[width=15.7cm]{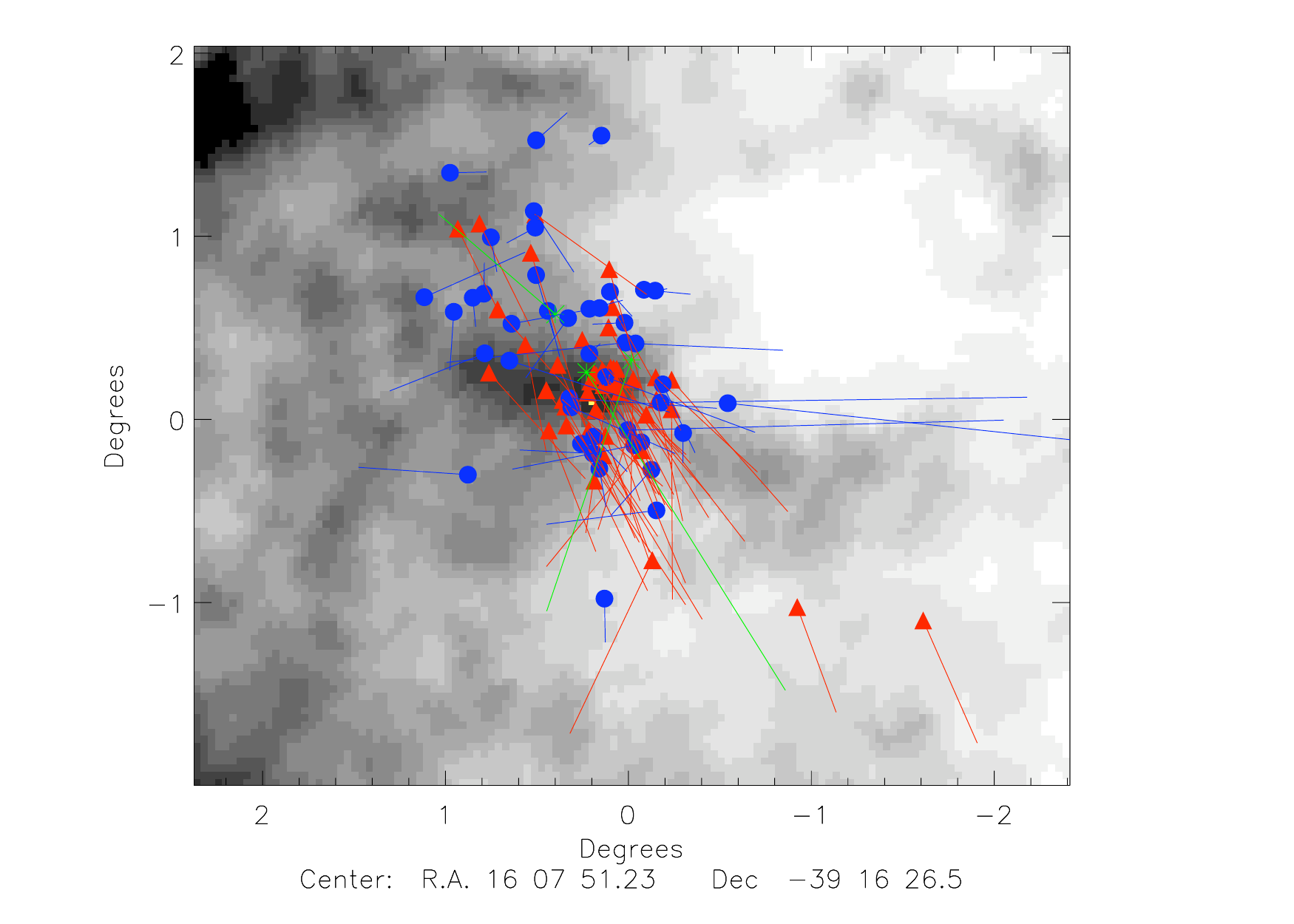}\hfill
       \caption{\footnotesize
	       Same as Fig.~\ref{fig:lup1dist} for Lupus~3.
	       }
         \label{fig:lup3dist}
   \end{figure*}

   \begin{figure*}[ht]
   \centering
    \includegraphics[width=15.7cm]{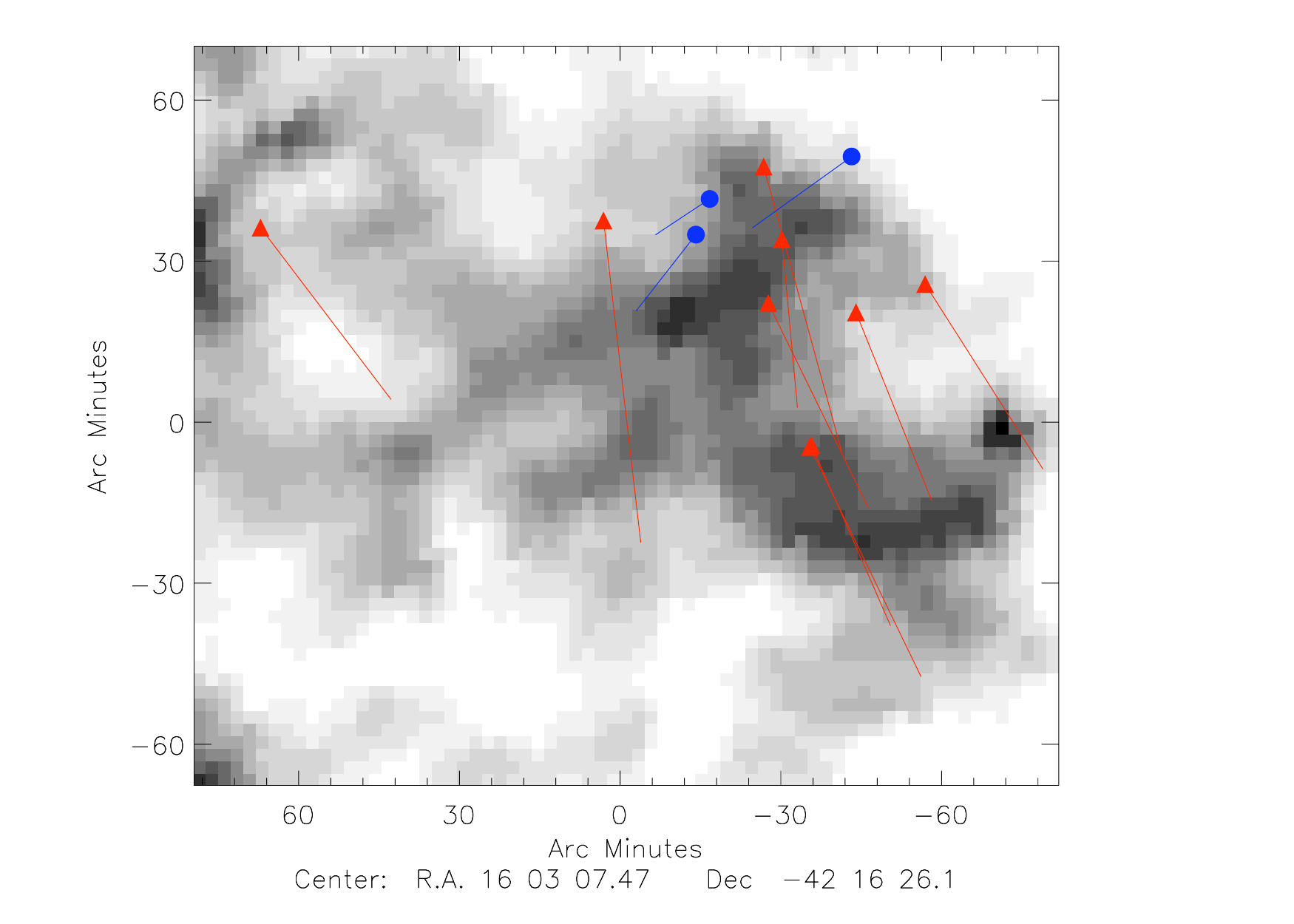}\hfill
       \caption{\footnotesize
	       Same as Fig.~\ref{fig:lup1dist} for Lupus~4.
	       }
         \label{fig:lup4dist}
   \end{figure*}

   \begin{figure}[t]
   \centering
  \includegraphics[width=8cm]{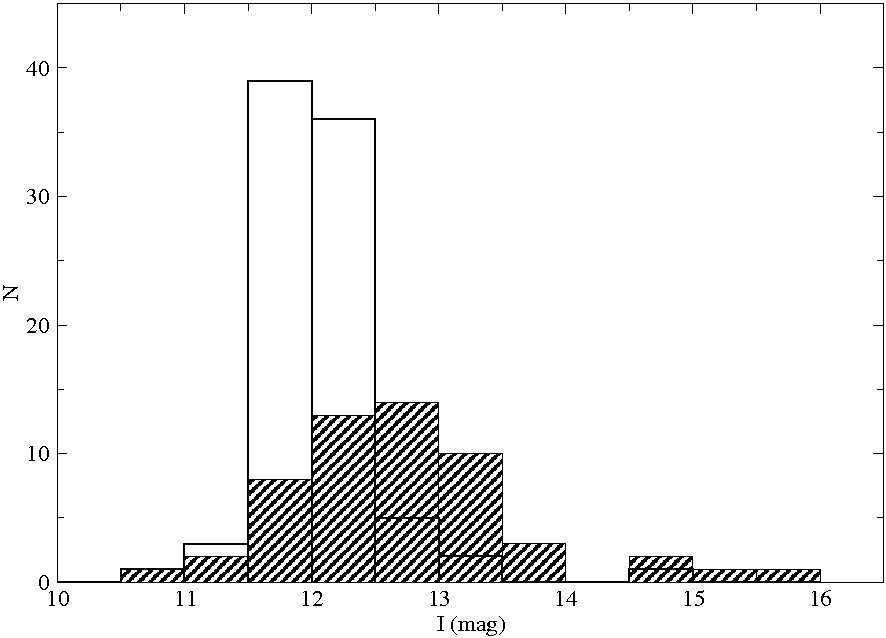}\hfill\\
  \includegraphics[width=8cm]{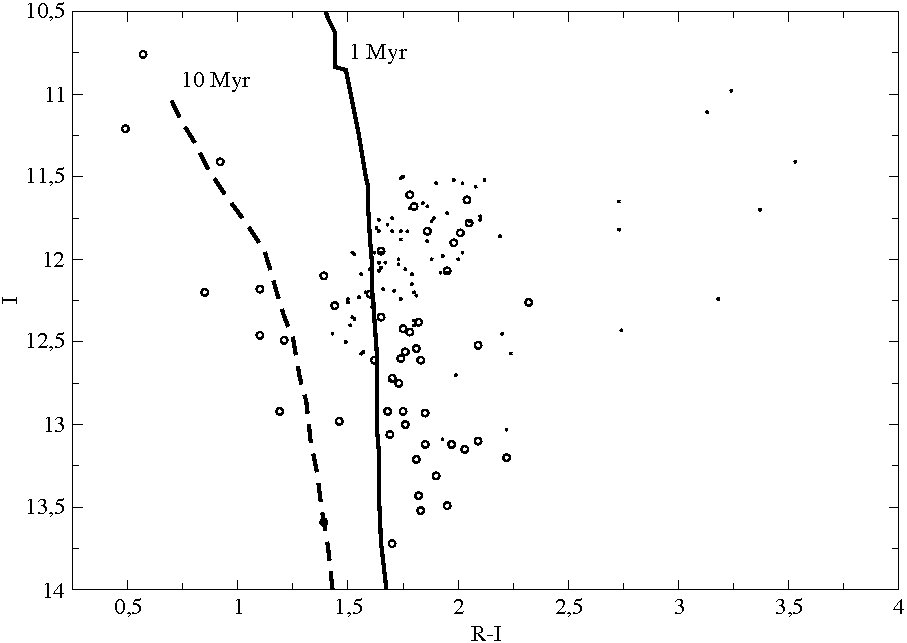}\hfill\\
  \includegraphics[width=8cm]{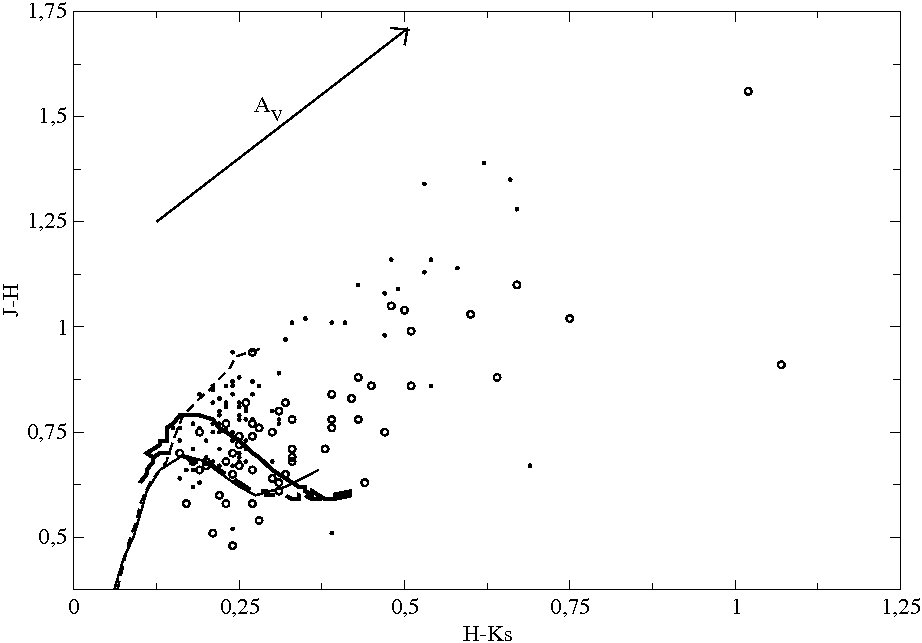}\hfill
      \caption{\footnotesize
      		Photometric properties of the Lupus members and candidate members.
	       {\em Upper panel:} $I$-magnitude distribution for Groups A and B (filled and blank histogram, respectively).
	       {\em Middle panel:}  ($I$, $R-I$) colour-magnitude diagram for the Lupus objects. Group A and Group B sources are indicated with open circles and dots, respectively. The thick solid and dashed lines are the 1 and 10~Myr isochrones, respectively, from the models by Baraffe et al. (\cite{baraffe98}).
                {\em Lower panel:}  ($J-H$, $H-Ks$) colour-colour diagram for the Lupus objects. The thin dashed and solid lines correspond to the giant and dwarf loci, respectively, from Bessell \& Brett (\cite{bessell88}). An extinction vector of 5~mag according to the law by Cardelli et al. (\cite{cardelli89}) is also indicated. Other symbols as in the previous panel. 
	       }
         \label{fig:phot}
   \end{figure}

Our Lupus candidate members are the YSOs in Lupus~1, 3, and 4 provided by  the  {\em c2d} program (Merin et al. \cite{merin08}) and the diskless cool stars identified in the optical WFI survey by Comer\'on et al. (\cite{comeron09}). We also incorporated the previously known members of these clouds from the recent review by Comer\'on (\cite{comeron08}). The three catalogues were cross-matched to avoid repeated entries and joined to produce a single catalogue for the complex. Table~\ref{tab:nobj} gives the number of objects provided by each work. The compiled lists included 90, 208, and 33 objects in Lupus~1, 3, and 4, respectively, making 331 in total. We note that most of these sources still lack spectroscopic confirmation of their youth and membership to the Lupus complex.

To get proper motion measurements of the Lupus candidate members, we cross-matched the merged catalogue with four astrometric catalogues that are available within the VO. These catalogues were the USNO-B1 (Monet et al. \cite{monet03}), the SuperCOSMOS Sky Survey (SSS) (Hambly et al \cite{hambly01}), the Positions and Proper Motions-Extended (PPMX) catalogue (R\"oser et al. \cite{roser08}), and the Third US Naval Observatory CCD Astrograph Catalog (UCAC3) (Zacharias et al. \cite{zacharias10}). We took advantage of VO tools for this task. The VO is a project designed to provide the astronomical community with the data access and the research tools necessary to enable the exploration of the digital, multi-wavelength universe resident in the astronomical data archives. In particular, to perform our cross-match, we made use of the multiple cone search utility of TOPCAT\footnote{\tt http://www.star.bris.ac.uk/$\sim$mbt/topcat/}, an interactive graphical viewer and editor for tabular data that allows the user to examine, analyse, combine, and edit astronomical tables. 

A matching radius of 2$\arcsec$ was used for all the catalogues, chosen after checking different values (1$\arcsec$, 2$\arcsec$, 3$\arcsec$, 10$\arcsec$) and finding 0.7$\arcsec$ and 0.4$\arcsec$ for the average separation and standard deviation, respectively, in the worst case.  After this step, we retained all sources whose proper motion errors were not set to zero, which had been computed using more than two epoch positions, and which in the case of UCAC3 had an object classification flag (ot) lower than 3, meaning it is a good star. After purging the data, the number of counterparts with proper motion data in USNO-B1, SSS, PPMX, and UCAC3 were 93, 266, 51, and 178, respectively. Table~\ref{tab:nobj} gives the number of objects with valid counterparts in each Lupus cloud. 

We compared the results of the cross-match with each catalogue. It is not possible to combine the data for the different astrometric catalogues because not all catalogues use the same reference system. After the comparison, we decided to use only the data provided by UCAC3 in the general analysis, for the following reasons: 1) USNO-B1 did not provide any proper motions lower than 6\,mas/yr; 2) due to its lower sensitivity, the number of counterparts found in PPMX is very low compared to UCAC3 or SSS; 3) in the case of SSS, although it provided the largest number of counterparts, the typical error of the proper motions is about twice the typical error of UCAC3. Thus, UCAC3 provided us with a large data set, with a suitable proper motion coverage and reasonable errors for our study. Therefore, only the UCAC3 data will be considered in the following.

The UCAC3 catalogue is an all-sky survey containing about 100 million objects, 95\% of them with proper motions, covering a dynamical range of 8-16~mag in the $R$ band. Its positional accuracy is about 15 to 100 mas per coordinate, depending on magnitude. The proper motion errors range from 1 to 10 mas/yr, depending on magnitude and observing history; however, some systematic errors may be present in the proper motions of stars with $R\ge14$. To make sure that the conclusions of this work were not significantly affected by the large errors of some of the faintest sources, we repeated the analysis described in the following section by using only the UCAC3 objects with $R<14$. The results using only this brighter sample showed excellent agreement with those obtained with the general sample. For this reason, and to provide better statistics, we decided to keep the faintest sources in our analysis, despite their larger errors.

Our final sample has 178 sources, the vast majority of them in Lupus~1 and 3. They correspond to 77\%, 46\%, and 39\% of the total number of sources included in our compiled lists for Lupus~1, 3, and 4, respectively.

%
\section{Kinematic groups towards the Lupus clouds}\label{sec:groups}

The left panel of Fig.~\ref{fig:pm} shows the vector point diagram for the sources included in the UCAC3 astrometric catalogue.  We clearly see two distinct groups, which are associated to the two peaks seen in the proper motion distribution (right panel): The first group  (``Group A'' in the following) is formed by sources with higher proper motions, where the mean value is around $\mu\sim28$~mas/yr. The objects belonging to the second group (``Group B'') have lower proper motions, with a mean close to $\mu\sim8$~mas/yr. 

For the analysis performed in the following sections, we define these two kinematic groups according to their total proper motion and their proper motion in the declination component, as they overlap a lot in $\mu_{\alpha}\cos\delta$. From the inspection of the corresponding histograms, we set the division between both groups at $\mu_{\delta}\sim-13.5$~mas/yr and $\mu\sim15$~mas/yr, where the two distributions reach a local minimum between the peaks corresponding to both groups. Of course, these criteria are somewhat arbitrary, and some sources are not clear members of either of the groups. These ambiguous cases are labelled in Tables~\ref{tab:groupA} and \ref{tab:groupB}. We also identify a number of sources with clearly discrepant proper motions, which are discussed in the Appendix. 

We carefully checked the reliability of the two groups by making use of all the information provided by the UCAC3 catalogue. We did not find any photometric bias or any particular flag or warning suggesting that one of the groups could be spurious or have lower quality than the other. Therefore, according to the UCAC3 data both groups are equally reliable. Moreover, because both groups are also recognized in the three clouds separately (in Lupus~4 less clearly owing to the low number of sources), they do not seem to be related to any particular cloud, but rather to the Lupus sky area as a whole. We note, though, that the errors of the Group B sources are on average lower than those of the Group A stars, which is probably related to the brighter magnitudes and spatial location of these objects (see Sect.~\ref{sec:ori}).

As a further check, we created vector point diagrams using the data from USNO-B1, SSS, and PPMX. A detailed comparison of these diagrams, taking the different errors and sensitivities of the data sets into account, shows very good agreement.

The conclusion is that both groups are probably real, meaning that the Lupus members and candidate members can be divided into two main populations. We list the objects belonging to each of these groups in Tables~\ref{tab:groupA} and \ref{tab:groupB}. The nature of these populations and their relation to the Lupus dark clouds is discussed in Sect.~\ref{sec:ori} below.

%
\section{Analysis and discussion}\label{sec:analysis}

\subsection{Origin of the kinematic groups}\label{sec:ori}

To clarify the origin of the kinematic groups and their relation to the Lupus complex, we carefully analysed the properties of the sources included in Groups A and B. Figures~\ref{fig:lup1dist} to \ref{fig:lup4dist} show the current spatial location of the sources included in the UCAC3 catalogue and their expected displacement within 10$^5$~Myr. We see that most of the Group A sources have similar proper motion directions (towards the southwest), suggesting a common origin. The Group B sources, on the other hand, exhibit a variety of proper motion directions, in the majority of cases totally discrepant with the one followed by Group A. This suggests that Group B may be formed by populations from various origins.

In Lupus~3, we also see differences in the spatial distribution of the sources. The objects in Group A tend to be more concentrated towards the cloud core than the objects in Group B, which tend to avoid the densest areas of the cloud. Outside the core, however, the distributions of both groups are indistinguishable. These differences in spatial distribution are not evident, however, in either of the other two clouds.

As seen in Tables~\ref{tab:groupAphot} and \ref{tab:groupBphot}, most of the Lupus members confirmed spectroscopically belong to Group A, because  the majority of studies have focused on the cloud cores. The only exceptions are Par-Lup3-2 (M6) and Sz~105 (M4), both in Lupus~3, which seem to belong to Group B, although their locations in the vector point diagram are not very far from the Group A sources.\footnote{\footnotesize We note that Sz~105 has a known companion (Ghez et al. \cite{ghez97}) at $\sim10\arcsec$, which is also included in UCAC3 (source 100-192094). The average proper motion of both sources has values consistent with membership in Group B.}

It is also noteworthy that Group B seems to be composed almost exclusively of diskless stars (class III sources) according to the literature (Mer\'{\i}n et al. \cite{merin08}; Comer\'on et al. \cite{comeron09}; see Table~\ref{tab:groupBphot}). The only exceptions are 160708.6-394723 and 161148.7-381758, which seem to harbour accretion disks (i.e., they are class II sources). In contrast, Group A includes stars both with and without disks (see Table~\ref{tab:groupAphot}). This suggests that Group B, as a whole, represents a later evolutionary stage than Group A, since disks are expected to be found around newly formed stars and to dissipate with time.

The optical and near-infrared photometry  from the literature (Mer\'{\i}n et al. \cite{merin08}; Comer\'on et al. \cite{comeron09}), also compiled in Tables~\ref{tab:groupAphot} and \ref{tab:groupBphot}, can help to clarify the nature of Group B. As shown in the upper panel of Fig.~\ref{fig:phot}, this group represents, as a whole, a brighter sample than Group A, which includes a significant amount of sources with $I>12$~mag. However, for similar $I$ magnitudes, the Group B stars tend to have redder $R-I$ colours, as seen in the ($I$, $R-I$) colour-magnitude diagram (central panel of Fig.~\ref{fig:phot}), which suggests higher extinction. Finally, in the near-infrared colour-colour diagram (lower panel of Fig.~\ref{fig:phot}) we see that, with a couple of exceptions, the Group B sources tend to be located along the reddening band for giants, although this locus can also be occupied by reddened young stars of late K and early M spectral types. The Group A sources, on the other hand, have colours that are more consistent with reddened mid-to-late M-dwarfs or with young stars with near-infrared excess. The brighter magnitudes and the optical and near-infrared colours suggest that the Group B stars are, on average, more massive and more extincted than the stars in Group A.

To conclude, all these facts (lower proper motions, different proper motion directions,  virtual lack of sources with disks, optical and near-infrared colours) strongly suggest that Group B represents a population (or a mixture of populations) located behind the dark clouds, which is probably unrelated to the complex. The true nature of these objects cannot be discerned with optical and infrared photometry alone and needs to be clarified with spectroscopy. Group A, on the other hand, has properties consistent with it being a genuine population of the Lupus star-forming region.

\subsection{Comparison to other populations}\label{sec:comp}

In an attempt to throw more light on the origin of the two kinematic groups identified towards the Lupus clouds, and on their relationship to larger structures, we compared their proper motions with those of other Gould Belt populations. As in the case of Lupus, we performed a multiple cone search in the UCAC3 catalogue to get proper motion measurements for these comparison objects. Figure~\ref{fig:pmcomp} shows again the vector point diagrams for the Lupus sources, where we have overplotted the objects from the populations discussed below.

\subsubsection{ROSAT pre-main sequence stars in the area of the Lupus clouds}\label{sec:rosat}

Krautter et al. (\cite{krautter96}) report the discovery by the ROSAT X-ray satellite of an extended population of weak-lined T~Tauri stars (WTTS) towards the Lupus clouds, which far exceeds in number the CTTS population coincident with the cloud cores. Some of these WTTS are seen projected onto regions of high obscuration, while others are located far from the cloud cores. In addition, the estimated age of this dispersed population is older than for the previously known Lupus stellar members. Wichmann et al. (\cite{wichmann97b}) argue that the age (up to 5-6\,10$^7$~yr) and spatial distribution of the ROSAT stars are consistent with their belonging to the Gould Belt rather than to the Lupus clouds themselves, whose mean age is estimated to be around 5~Myr. 

The UCAC3 catalogue contains counterparts for 91 of the 136 ROSAT stars from Krautter et al. (\cite{krautter96}). These sources are, with a few exceptions, located in the same area as Group A in the vector-point diagram of Fig.~\ref{fig:pmcomp}, a fact that suggests that both populations may have a similar origin. Only six ROSAT stars are seen detached from this group, and a couple of them, namely RXJ1506.9-3714 and RXJ1605.5-3846, are located in the Group B area.

\subsubsection{Members of the Upper Scorpius association}\label{sec:usco}

The Upper Scorpius (USco) OB association belongs to the largest Scorpius-Centaurus structure within the Gould Belt. Thus, it is located not far from the Lupus clouds. Over the past decade, it has been one of the most targeted star-forming regions for the search and study of low-mass stars and very low-mass objects. Combining samples of USco members from the literature, Bouy \& Mart\'{\i}n (\cite{bouy09}) have recently performed a kinematic study of the low-mass population in this region using the USNO-B1 and UCAC2 catalogues. We performed the multiple cone search in the UCAC3 catalogue using the same compiled list as these authors, kindly provided to us by H. Bouy. It is important to note that all the sources studied by Bouy \& Mart\'{\i}n (\cite{bouy09}) have been spectroscopically confirmed as young low-mass stars.

\begin{figure*}[ht]
   \centering
  \includegraphics[width=16cm]{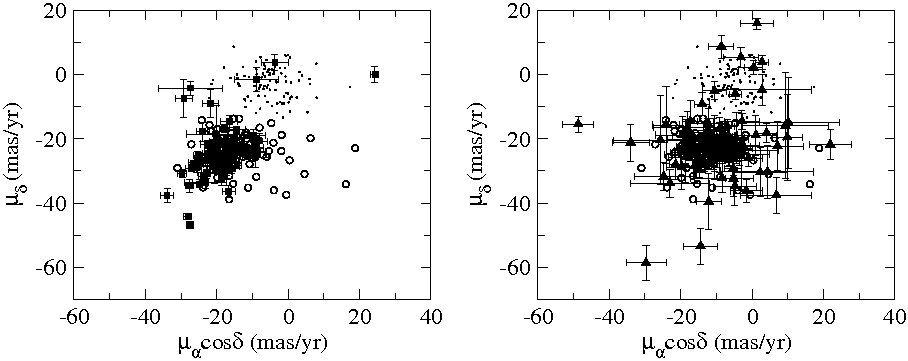}\hfill
      \caption{\footnotesize
	      Comparison of proper motion components of the Groups A and B sources (open circles and dots, respectively) with those of other Gould Belt populations: Lupus ROSAT stars (filled squares, left panel) and Upper Scorpius members (filled triangles, right panel). }
         \label{fig:pmcomp}
   \end{figure*}

The USco UCAC3 sample contains 129 of the 514 objects in the list of Bouy \& Mart\'{\i}n. In Fig.~\ref{fig:pmcomp} we see that  the majority of them are coincident in proper motion with our Group A sources, again suggesting a similar origin. Nonetheless, nine stars have proper motions that seem to agree with Group B. Of them, one star, USco~J161026.4-193950, has been identified as a kinematic member of Upper Scorpius by Bouy \& Mart\'{\i}n (\cite{bouy09}). Another source, USco~J155744.9-222351, is classified as an outlier by these authors.  The remaining objects (UScoCTIO~31, UScoCTIO~36, USco~J160132.9-224231, USco~J160545.4-202308, and USco~J161021.5-194132) do not have available proper motions in the work by Bouy \& Mart\'{\i}n (\cite{bouy09}). 

\subsubsection*{}\label{}

In summary, the proper motions of the Group A sources fully agree with those of other Gould Belt populations, which reinforces the idea that this group, and hence the Lupus dark cloud complex, is associated to this large structure. A few stars from other Gould Belt structures have similar proper motions to those of the Group B sources; however, they are very few, and without further information (e.g. radial velocities), we cannot rule out that this is just a matter of chance due to the projection of their true spatial motions on the sky.

\subsection{Properties of the Lupus kinematic members}\label{sec:prop}

Our kinematic analysis has shown that the sources in Group A are likely to be members of the Lupus star-forming region, while Group B seems to represent a background population. In particular, about one third of the Group A sources have been spectroscopically confirmed as pre-main sequence stars. The definitive confirmation of youth for the rest of the objects needs to await for spectroscopy; nonetheless, because of the good agreement of their proper motions with those of the confirmed members, we expect contamination to be minimal. 

In the following discussion, we thus consider all the stars in Group A, and only the stars in that group, as probable Lupus members.  This group contains 75 stars, of which 25 have spectroscopic confirmation of their status. The embedded and/or highly extincted population, as well as the lowest mass members of the Lupus complex ($M<0.1M_{\odot}$ ), are certainly missing in this sample, due to the relatively bright magnitude limit of the astrometric catalogue.

\subsubsection{Spatial location}\label{sec:dist}

As shown in Figs.~\ref{fig:lup1dist} to \ref{fig:lup4dist}, the Group A sources (our Lupus kinematic members) exhibit similar proper motions in all three clouds, as expected if these clouds belong to the same large structure. Moreover, these sources can be divided into two samples according to their spatial location: a first, more numerous group of objects that are seen concentrated towards the cloud cores, and a second group of stars which are dispersed around the densest cloud areas. However, no evident difference is seen between the in- and the off-cloud population, neither in proper motion moduli nor in the proper motion directions. This seems to confirm that both samples belong to the same structure. It also suggests that the location of the outer sources is not the consequence of these objects being ejected from their parental birth sites.

\begin{figure*}[t]
   \centering
  \includegraphics[width=16cm]{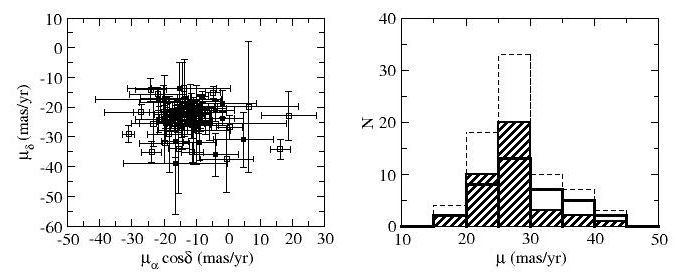}\hfill
      \caption{\footnotesize
	       {\em Left panel:} Vector point diagrams for the Lupus kinematic members (our Group A). Class I/II sources are marked with filled circles, class III sources with open squares. 
	       {\em Right panel:} Total proper motion distribution for the whole sample (dashed histogram) and for the class I/II and class III sources separately (hashed and blank histogram, respectively).}
         \label{fig:pmclass}
   \end{figure*}

Two stars (Sz~92 and SSTc2d~J160734.3-392742) seem to be moving differently in Fig.~\ref{fig:lup3dist}, as though they were escaping from the Lupus~3 cloud. These sources are located on the right edge of the Lupus group in the vector point diagram, slightly detached from the rest (see left panel of Fig.~\ref{fig:pmclass}). We could restrict our selection in the vector point diagram to avoid these runaway sources; however, that one of them (Sz~92) has been confirmed spectroscopically as a member of Lupus suggests that their discrepant proper motions may have a different cause than non-membership. Dynamical interactions with other cloud members or unresolved binary companions may be responsible for the measured discrepant proper motions. Given the large errors ($\sim3$ and $\sim8.5$~mas/yr, respectively), it is also possible that more accurate  measurements are able to reconcile their proper motions with those of the majority of Lupus members.

\subsubsection{Binarity}\label{sec:bin}

Since the presence of unresolved binaries in the images used for the construction of the UCAC3 catalogue can affect the proper motion measurements, we checked the binarity of our probable Lupus members. Our sample includes 12 reported visual binaries and one triple system, Sz~130  (Reipurth \& Zinnecker \cite{reipurth93}; Ghez et al. \cite{ghez97}; Mer\'{\i}n et al. \cite{merin08}). They are labelled in Table~\ref{tab:groupA}. We do not find any relation between the presence of a companion and the proper motion modulus or direction of the system; in other words, not all the discrepant objects are known multiple systems, nor do all known multiple systems have discrepant proper motions. 

We note that the separation of the reported companions is relatively large (between 1 and 10$\arcsec$, except for Sz~74, whose companion is located at 0.24$\arcsec$), as most of them have been identified in visual searches. To date, the vast majority of companions still lack confirmation of their youth and membership in the Lupus clouds. We tried to throw more light on this issue by looking for proper motion measurements for these companions in UCAC3. However, due to their faintness, they are not included in this catalogue. The only exception is the reported companion to Sz~91 in Lupus~3, for which we find a possible counterpart at 8.3$\arcsec$ to the South-East (UCAC3~102-194472, $\mu_{\alpha}\cos\delta=12.1$~mas/yr, $\mu_{\delta}=-40.6$~mas/yr). The proper motion of the combined system looks compatible with its belonging to the Lupus kinematic group.

\subsubsection{Stars with and without disks}\label{sec:disks}

   \begin{figure*}[t]
   \centering
  \includegraphics[width=15.5cm]{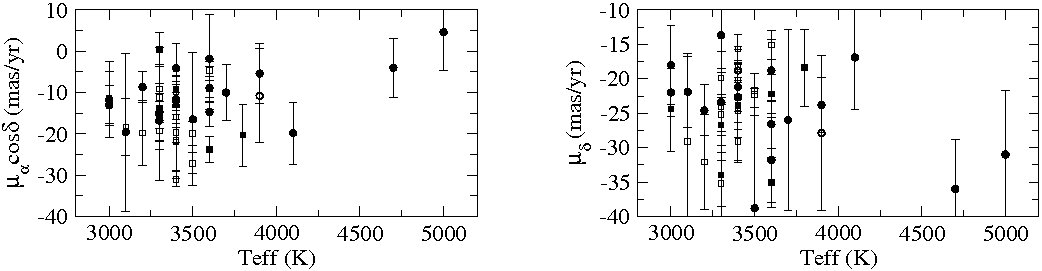}\hfill     
 \caption{\footnotesize
	       Proper motion components of our Lupus kinematic members plotted against effective temperature. Solid and open symbols stand for objects with and without spectroscopic confirmation of youth, respectively. Class II and class III sources are indicated with circles and squares, respectively. }
         \label{fig:pmteff}
   \end{figure*}
 
   \begin{figure*}[t]
   \centering
  \includegraphics[width=16cm]{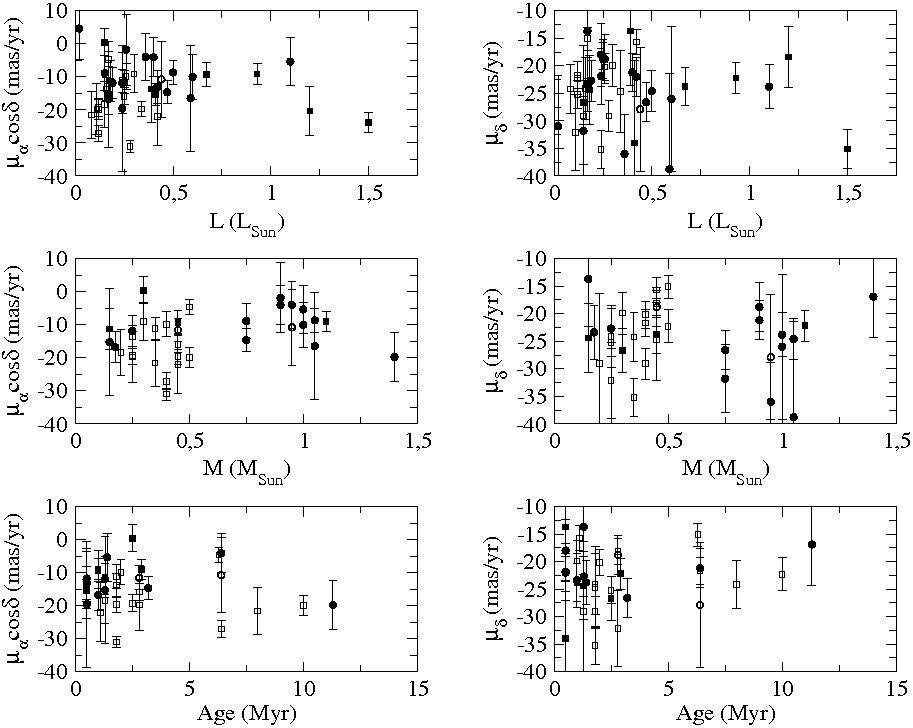}\hfill     
      \caption{\footnotesize
	       Proper motion components of the Lupus kinematic members plotted against luminosity, mass, and age of the objects (upper, middle, and lower panels, respectively). Symbols as in Fig.~\ref{fig:pmteff}}
         \label{fig:pmpars}
   \end{figure*}

According to the analysis of their spectral energy distributions (SEDs) performed by Mer\'{\i}n et al. (\cite{merin08}) and Comer\'on et al. (\cite{comeron09}), our UCAC3 sample contains 36 stars with disks (class II sources) and 37 diskless stars (class III sources), as well as one star thought to possess a disk and a circumstellar envelope (a class I source), namely Sz~102 in Lupus~3. Another source (160836.2-392302, also in Lupus~3) could be in an intermediate state between classes I and II (Comer\'on et al. \cite{comeron09}), but it will be considered a class II source in the following. The SED classes of our Group A sources are indicated in Table~\ref{tab:groupAphot}.

According to these numbers, the disk fraction in our kinematic member list is about 50\%, lower than the 70-80\% estimated by Mer\'{\i}n et al. (\cite{merin08}) in their $c2d$ $Spitzer$ survey. Although this could suggest that our sample is incomplete in terms of disks, we note that a significant fraction of the diskless stars in our sample come from the survey by  Comer\'on et al. (\cite{comeron09}), which was published after the Mer\'{\i}n et al. study.  Our estimated disk fraction is, in any case, within the range observed in other star-forming regions of similar ages (Haisch et al. \cite{haisch01}; Mamajek \cite{mamajek09}), so we believe it is quite representative of the disk composition of the Lupus population down to about 0.1$M_{\odot}$ (corresponding to $I\sim14.5$~mag at the age and distance of the Lupus clouds, according to Baraffe et al. \cite{baraffe98}). 
 
We searched for correlations between the existence of a disk and the proper motion properties of the objects. In Fig.~\ref{fig:pmclass}, we show the vector-point diagram and the total proper motion histograms for the objects in our member list, divided by SED class (class I and II versus class III). The proper motions of class I-II and class III are indistinguishable with the available data.

As already reported by other authors (e.g. Hughes et al. \cite{hughes94}; Mer\'{\i}n et al. \cite{merin08}; Comer\'on et al. \cite{comeron09}), most of the objects located outside the cloud cores are diskless stars, and thus theoretically more evolved than the in-cloud population, which consists mainly of stars with disks. However, the estimated ages of both groups are similar (Comer\'on et al. \cite{comeron09}). A possible explanation for the apparently faster evolution of the off-cloud population is that these sources had been ejected from their parental birth sites, hence devoid of circumstellar material to accrete. However, as stated in Sect.~\ref{sec:dist} above, that the proper motions of these sources are  very similar to those of the in-cloud population seems to contradict this picture. Comer\'on et al. (\cite{comeron09}) propose that the off-cloud population formed after the passage of one or more shock fronts associated with evolved supernova remnants, triggering the collapse of pre-existing cloudlets. The outside-in collapse produced by this scenario, as opposed to the inside-out collapse of the cores embedded in the clouds, may explain the different disk properties of both populations. Another alternative explanation would be the dissipation of unshielded disks by the ultraviolet radiation of O-stars from the Scorpius-Centaurus OB association, but this mechanism is more doubtful given the distance of the association to the Lupus clouds.

\subsubsection{Spectral types and physical parameters}\label{sec:spt}

Spectral types are related to the effective temperature of the objects and, in the case of PMS stars, they are also directly related to the mass. Bouy \& Mart\'{\i}n (\cite{bouy09}) have investigated the relation between proper motion and spectral type for their sample of Upper Scorpius members, and found no correlation between these parameters. We now attempt to make a similar check with our Lupus kinematic members.

As mentioned previously, only 25 objects in our member list have spectral types available in the literature (see Table~\ref{tab:groupAphot}). Except for three K-type stars (Sz~102, Sz~133, and Sz~118), they span the range M0-M6, often with only two or three objects in each subclass. 

To increase our sample, we made use of effective temperatures instead of spectral types. The temperatures were estimated in two different ways. For those sources with published spectral types, we transformed the spectral types into temperatures using the scale by Luhman et al. (\cite{luhman03}). Since some authors (e.g. Barrado y Navascu\'es et al. \cite{byn04}) have found discrepancies of up to $200$~K between the Luhman scale and other temperature estimations, we consider an error of 200~K for these $T_{eff}$ values. For the rest of sources, we directly took the effective temperatures provided by  Comer\'on et al. (\cite{comeron09}; see their Tables 4 and 9), derived by fitting the SEDs of the objects using model spectra from Hauschildt et al. (\cite{hauschildt99}) and Allard et al. (\cite{allard00}). For $T_{eff}<4000$~K,  these authors again estimate an error of $\pm200$~K in their temperature values. We note that the Luhman scale was set to be compatible with the Baraffe et al. (\cite{baraffe98}) evolutionary models, which are based on the same atmospheric models that Comer\'on et al. used for their SED fitting. 

It is important to point out that Comer\'on et al. only attempted to fit the stellar photospheres of the objects, thus assuming any contribution from the disk or the accretion to the total flux to be negligible. As a consequence, their fits are less reliable for stars with strong accretion and/or high near-infrared excesses; indeed, Comer\'on et al. (\cite{comeron09}) caution that, in such cases, the derived effective temperatures and extinctions are too high. The number of stars in our member list with reliable fits in that work is 28 (twenty diskless stars and eight stars with disks), of which 10 have published spectral types. Hence, we have effective temperatures derived from the SED fitting for 18 objects, most of them class III sources. Together with the 25 stars with spectral types, this makes a sample of 43 sources. The estimated effective temperatures,  which are listed in 
Table~6, range between 5000 and 3000~K, with a majority of stars having $T_{eff}<4000$~K, as expected for M-type objects.

Figure~\ref{fig:pmteff} shows the proper motion components of our objects plotted against effective temperature. The plots show no correlation between these parameters. No clear trend is seen either when stars with and without disks are considered separately. 

Using the parameters from the SED fitting, Comer\'on et al. (\cite{comeron09}) derive absolute luminosities, masses, and ages for the objects from the evolutionary models by Baraffe et al. (\cite{baraffe98}) and Chabrier et al. (\cite{chabrier00}). They assumed a distance of 200~pc to the complex. The thus derived physical parameters for our Lupus kinematic members are also given in %
Table~6.

According to their results, the 28 stars with reliable fits in our member list have masses between 0.1 and 1.4$M_{\odot}$, luminosities in the range 0.08-1.1$L_{\odot}$ (except for one object, Sz~118, with $L=4.3L_{\odot}$) and estimated ages younger than 15~Myr.  We used this subsample to look for correlations between the proper motions and the physical parameters of the objects. However, as shown in Fig.~\ref{fig:pmpars}, no clear variations are seen in the proper motions with luminosity, mass, or age. 
More accurate proper motion measurements would be required to reveal any eventual trend with the physical properties of our objects.

%
\section{Conclusions}\label{sec:concl}

Our kinematical analysis of the Lupus members and candidate members has unveiled two main groups of sources. The first one, our Group A, has proper motions, spatial distribution, and colours consistent with it being a genuine population of the star-forming complex. Besides, we have shown that the proper motions of the sources in this group are consistent with those of other Gould Belt populations, thus reinforcing the idea of a physical connection between the dark clouds and this large structure. The second group, our Group B, has lower proper motions with random directions, and the sources in this group tend to avoid the location of the cloud cores. In addition, the optical and near-infrared colours of these objects are consistent with them being reddened giants. All this suggests that Group B represents a background population or mixture of populations, probably unrelated to the Lupus clouds. However, a few spectroscopically confirmed members of Lupus and of other Gould Belt structures have locations in the vector point diagram that are consistent with Group B, leaving open the possibility that this group may contain some other true Lupus members whose proper motions are, for some reason, discrepant with respect to the majority of objects belonging to the complex.

We investigated any possible relationship between the proper motions and other properties of these sources, such as spatial location, binarity, the presence of an accretion disk, and several physical parameters (effective temperature, luminosity, mass, age), but we have found no evident trends with the available data. In particular, we do not see any differences between the objects located within the cloud cores and those located outside these cores. This suggests that the dispersed population has not been ejected from the densest cloud areas, but has actually formed outside the cores. More accurate proper motions, hopefully combined with radial velocity information, can unveil any possible hidden trend undetectable with the precision of the proper motions used here, thus helping throw more light on the formation process of the stellar content of the Lupus clouds.

Our study has shown how the use of kinematical information can complement photometric data to constrain the selection of members of young star-forming clusters, minimizing the contamination of the samples. Currently available astrometric catalogues are deep enough to provide proper motions for a significant number of candidate members of nearby star-forming regions. Kinematic information can also help towards a better understanding of the formation process of these objects, especially if it can be combined with physical parameters derived from SED fitting and/or spectroscopic data.  The {\em Gaia} mission of the European Space Agency, whose launch is foreseen for 2012, will provide this information with unprecedented precision for most of the stars in the sample studied here, allowing a three-dimensional study of Lupus and other nearby star-forming regions.

 \begin{acknowledgements}
 
We kindly thank H. Bouy, F. Comer\'on, B. Mer\'{\i}n and J.~A. Caballero for useful discussions. 

This work was funded by the Spanish MICINN through grant Consolider-CSD2006-00070. It also benefited from funding from the Spanish government through grants ESP2007-65475-C02-02, AYA2010-21161-C02-02 and from the Madrid regional government through grant PRICIT-S2009ESP-1496. This research has made use of the Spanish Virtual Observatory supported from the Spanish MEC through grant AYA2008-02156.
 
This publication greatly benefited from the use of the SIMBAD database and VIZIER catalogue service, both operated at the CDS (Strasbourg, France). We used the VO-compliant tools Aladin, developed at the CDS, and TOPCAT, currently developed within the AstroGrid project. 

\end{acknowledgements}

%
\begin{appendix}
\section{Outlier analysis}\label{sec:out}

\begin{table*}[ht]
\label{tab:outl}
\caption{Outliers in the Lupus vector point diagram}
\centering
\begin{tabular}{c c c c c r r c}
\hline
\hline\noalign{\smallskip}
  \multicolumn{1}{c}{SSTc2d~J} &
   \multicolumn{1}{c}{Other names} &
   \multicolumn{1}{c}{$\alpha$ (J2000)} & 
    \multicolumn{1}{c}{$\delta$ (J2000)}  &
   \multicolumn{1}{c}{UCAC3} &
 \multicolumn{1}{c}{$\mu_{\alpha}\cos\delta$} &
  \multicolumn{1}{c}{$\mu_{\delta}$} &
   \multicolumn{1}{c}{Notes} \\
  \multicolumn{1}{l}{} &
   \multicolumn{1}{c}{} &
   \multicolumn{1}{c}{} &
   \multicolumn{1}{c}{} &
   \multicolumn{1}{c}{} &
 \multicolumn{1}{c}{(mas/yr)} &
  \multicolumn{1}{c}{(mas/yr)} &
   \multicolumn{1}{c}{} \\
\noalign{\smallskip}\hline\noalign{\smallskip}
  \multicolumn{8}{c}{Lupus 1 cloud}\\
  \noalign{\smallskip}
  153928.3-344618 &              & 15:39:28.3 & $-$34:46:18 & 111-179768 & $157.6\pm5.6$ & $-40.2\pm5.6$ & 1 \\
  153940.8-333941 &              & 15:39:40.8 & $-$33:39:41 & 113-182897 & $-16.0\pm 3.7$ & $-65.0\pm2.5$ & \\
  154144.0-343530 &              & 15:41:44.0 & $-$34:35:30 & 111-180079 & $-255.7\pm10.2$ & $-87.8\pm9.8$ & \\
  154344.5-335834 &              & 15:43:44.5 & $-$33:58:34 & 113-183752 & $-43.9\pm1.6$ & $-11.5\pm1.6$ & \\
\noalign{\smallskip}\hline\noalign{\smallskip}
    \multicolumn{8}{c}{Lupus 3 cloud}\\
    \noalign{\smallskip}
  160539.8-384811 &              & 16:05:39.8 & $-$38:48:11 & 103-200987 &  168.8 $\pm$  6.9 & -27.3 $\pm$  7.5 & 1, 2 \\  
  160752.3-385806 & Sz~95 & 16:07:52.3 & $-$38:58:06 & 103-202637  & $16.8\pm8.4$ & $-49.4\pm15.3$ & \\
  160908.5-390213 &              & 16:09:08.5 & $-$39:02:13 & 102-194972 & $-40.2\pm8.3$ & $-62.1\pm3.2$ & \\
  161001.1-384315 &              & 16:10:01.1& $-$38:43:15 & 103-203772 & $22.7\pm2.9$ & $19.8\pm2.8$ & \\
  \noalign{\smallskip}\hline\noalign{\smallskip}
    \multicolumn{8}{c}{Lupus 4 cloud}\\
    \noalign{\smallskip}
  160111.6-413730 &             & 16:01:11.6 & $-$41:37:30 & 097-208263 & $76.9\pm28.2$ &  $50.0\pm29.5$ & 1 \\
  \hline
\end{tabular}
\begin{flushleft}
{\bf Notes.} \\
     (1) 
       The UCAC3 proper motion does not agree with the measured in the archive images (see text). \\
     (2) This source has a reported visual companion without UCAC3 counterpart.
\end{flushleft}
\end{table*}

Some objects present proper motions that are clearly discrepant from those of the stars in Groups A and B. In this Appendix, we  discuss the reliability of the proper motion measurements for these objects, as listed in Table~\ref{tab:outl}.

Initially, nine sources were identified as outliers in the vector point diagram: four in Lupus~1, four in Lupus~3, and one more in Lupus~4. To check the reliability of the reported proper motions, we visually inspected these objects with the peculiar proper motions using Aladin. We compared and blinked two sets of images separated several decades in time, from the optical DSS-1 and the near-infrared 2MASS surveys, to easily confirm the high proper motion of our candidates. The 2MASS (Skrutskie et al. \cite{skrutskie06}) sources and UCAC3 counterparts were superimposed on the images to assess the reliability of the cross-match. Besides, we used other available astro-photometric databases, such as the Astrographic Catalogue AC2000.2 (Urban et al. \cite{urban98}), the SuperCOSMOS Sky Survey, the USNO-B Catalog and the PPMX catalogue, to verify the peculiar proper motions.

From the original group of nine sources, six turned out to have reliable proper motions that are discrepant from those of Groups A and B. Figures~\ref{fig:lup1dist} to \ref{fig:lup4dist} show their location and their expected displacement after 10$^5$~yr. Although most of the outliers are located outside the densest cloud areas, two stars (153940.8-333941 in Lupus~1 and 160908.5-390213 in Lupus~3) are seen towards the cloud cores. The objects from Table~\ref{tab:outl} with wrong proper motions are not plotted in these figures.

The proper motion values and directions differ widely among these sources. Three objects are seen not far, but clearly detached, from Groups A and B in the vector point diagram, having $\mu\la50$~mas/yr: 154344.5-335834 in Lupus~1 and 161001.1-384315 and Sz~95 in Lupus~3. We note that the latter has a reported visual companion at a separation of $\sim3\arcsec$ according to Comer\'on (\cite{comeron08}), but this second object has no UCAC3 counterpart.

The remaining sources in the list of outliers have $50<\mu<100$~mas/yr, except for 154144.0-343530 in Lupus~1, which has an even higher value, $\mu\sim270$~mas/yr. Such high proper motions suggest they might be foreground sources.

With a single exception (Sz~95), all these outliers are diskless stars according to the SED analysis (Mer\'{\i}n et al. \cite{merin08}; Comer\'on et al. \cite{comeron09}), a fact that is consistent with them belonging to the field. However, ejected members from star-forming regions are expected to possess only truncated disks, which would thus dissipate much faster than disks around the in-cloud population. Should some of them be members of the Lupus clouds, these objects would have masses around 0.3-0.4$M_{\odot}$ according to the SED-fitting results by Comer\'on et al. (\cite{comeron09}), very close to the median mass of the stars in the region. 

\end{appendix}




\addtocounter{table}{1}  
\longtabL{2}{
\begin{longtable}{cc cc c rr c}

\caption{~~~~~~~~~~~~~~~~~~~~~~Table~\ref{tab:groupA}: Group A sources (Lupus kinematic members)}
\label{tab:groupA}\\
    \hline \hline \noalign{\smallskip}
  \multicolumn{1}{c}{SSTc2d~J} &
   \multicolumn{1}{c}{Other names} &
   \multicolumn{1}{c}{$\alpha$ (J2000)} & 
    \multicolumn{1}{c}{$\delta$ (J2000)}  &
   \multicolumn{1}{c}{UCAC3} &
 \multicolumn{1}{c}{$\mu_{\alpha}\cos\delta$} &
  \multicolumn{1}{c}{$\mu_{\delta}$} &
   \multicolumn{1}{c}{Notes} \\
  \multicolumn{1}{l}{} &
   \multicolumn{1}{c}{} &
   \multicolumn{1}{c}{} &
   \multicolumn{1}{c}{} &
   \multicolumn{1}{c}{} &
 \multicolumn{1}{c}{(mas/yr)} &
  \multicolumn{1}{c}{(mas/yr)} &
   \multicolumn{1}{c}{} \\
      \noalign{\smallskip}\hline\noalign{\smallskip}
\endfirsthead

\caption{~~~~~~~~~~~~~~~~~~~~~~Table~\ref{tab:groupA}: continued.}\\
   \hline\hline\noalign{\smallskip}
  \multicolumn{1}{c}{SSTc2d~J} &
   \multicolumn{1}{c}{Other names} &
   \multicolumn{1}{c}{$\alpha$ (J2000)} & 
    \multicolumn{1}{c}{$\delta$ (J2000)}  &
   \multicolumn{1}{c}{UCAC3} &
 \multicolumn{1}{c}{$\mu_{\alpha}\cos\delta$} &
  \multicolumn{1}{c}{$\mu_{\delta}$} &
   \multicolumn{1}{c}{Notes} \\
  \multicolumn{1}{l}{} &
   \multicolumn{1}{c}{} &
   \multicolumn{1}{c}{} &
   \multicolumn{1}{c}{} &
   \multicolumn{1}{c}{} &
 \multicolumn{1}{c}{(mas/yr)} &
  \multicolumn{1}{c}{(mas/yr)} &
   \multicolumn{1}{c}{} \\
    \noalign{\smallskip}\hline\noalign{\smallskip}
\endhead
\hline
\endfoot

  \multicolumn{8}{c}{Lupus~1 Cloud}\\
  \noalign{\smallskip}
  153805.7-341535 &  	        & 15:38:05.7 & -34:15:35 & 112-181227 &  $ -5.6 \pm  1.5$ & $ -3.3 \pm  1.4$ & \\
  153905.2-341210 &  	        & 15:39:05.2 & -34:12:10 & 112-181437 &  $ -8.3 \pm  2.1$ & $ -6.6 \pm  2.3$ & \\
  153927.8-344617 &  Sz~65, IK~Lup & 15:39:27.9 & -34:46:17 & 111-179766 &  $-21.7 \pm 19.6$ & $-17.5 \pm  3.5$ & \\
  154009.4-342734 &                & 15:40:09.4 & -34:27:34 & 112-181631 &  $-31.0 \pm  1.8$ & $-29.1 \pm  2.8$ & \\
  154013.7-340142 &                & 15:40:13.7 & -34:01:42 & 112-181650 &  $ -4.7 \pm  2.3$ & $-15.1 \pm  2.1$ & 1 \\
  154018.5-342614 &                & 15:40:18.5 & -34:26:14 & 112-181670 &  $-19.3 \pm  2.6$ & $-25.2 \pm  2.5$ & \\
  154122.0-344015 &                 & 15:41:22.0 & -34:40:15 & 111-180007 &  $-27.1 \pm  2.5$ & $-21.7 \pm  2.5$ & \\
  154148.3-350145 &                 & 15:41:48.3 & -35:01:45 & 110-180365 &  $-19.9 \pm  3.0$ & $-22.3 \pm  3.0$ & \\
  154201.8-345449 &                 & 15:42:01.8 & -34:54:49 & 111-180129 &  $-17.9 \pm  2.2$ & $  2.2 \pm  2.2$ & \\
  154257.6-341928 &  	         & 15:42:57.6 & -34:19:28 & 112-182163 &  $-24.2 \pm  4.8$ & $-14.2 \pm  3.9$ & \\
  154344.5-335834 &  	         & 15:43:44.5 & -33:58:34 & 113-183752 &  $-43.9 \pm  1.6$ & $-11.5 \pm  1.6$ & \\
  154512.9-341731 & Sz~68, HT~Lup  & 15:45:12.9 & -34:17:31 & 112-182463 &  $-15.7 \pm  1.8$ & $-20.7 \pm  1.2$ & 2 \\
  154517.4-341829 & Sz~69, HW~Lup & 15:45:17.4 & -34:18:28 & 112-182467 &  $-10.0 \pm  6.8$ & $-26.0 \pm 13.2$ & 2 \\
  	                          & Sz~70      & 15:46:43.0 & -34:30:12 & 111-181035 &  $-15.8 \pm  4.5$ & $-21.5 \pm  3.9$ & \\
  	                          & Sz~71      & 15:46:44.7 & -34:30:35 & 111-181040 &  $ -1.9 \pm  6.5$ & $-23.8 \pm  2.5$ & \\
  	                          & Sz~72      & 15:47:50.6 & -35:28:35 & 110-181840 &  $-15.2 \pm  2.5$ & $-25.3 \pm  2.5$ & \\
  	                          & Sz~73      & 15:47:56.9 & -35:14:35 & 110-181858 &  $-16.5 \pm  4.2$ & $-31.5 \pm  5.3$ & \\
  	                          & Sz~74      & 15:48:05.2 & -35:15:53 & 110-181875 &  $ -6.0 \pm 24.0$ & $-25.6 \pm  5.5$ & 2 \\
  	                          & Sz~75      & 15:49:12.1 & -35:39:04 & 109-183258 &  $-15.1 \pm  2.8$ & $-23.4 \pm  2.5$ & \\
  	                          & Sz~76      & 15:49:30.8 & -35:49:52 & 109-183338 &  $-15.9 \pm  2.0$ & $-26.4 \pm  2.0$ & \\
  	                          & Sz~77      & 15:51:47.0 & -35:56:43 & 109-183891 &  $-11.1 \pm  2.1$ & $-25.4 \pm  1.4$ & 2 \\

  \noalign{\smallskip}\hline\noalign{\smallskip}
  \multicolumn{8}{c}{Lupus~3 Cloud}\\
  \noalign{\smallskip}
                                   & RY~Lup  & 15:59:28.4 & -40:21:51 & 100-186413 &  $-11.4 \pm  1.3$ & $-23.7 \pm  1.2$ & \\ 
		                & EX~Lup   & 16:03:05.5 & -40:18:26 & 100-188355 &  $ -8.1 \pm  7.4$ & $-20.5 \pm  3.3$ & \\
                                  & Sz~86      & 16:06:44.3 & -39:14:11 & 102-194310 &  $ -0.5 \pm  8.2$ & $-37.4 \pm 11.4$ & \\
   160644.6-390432 &                & 16:06:44.6 & -39:04:32 & 102-194315 &  $-23.3 \pm  4.3$ & $-25.6 \pm  7.6$ & \\
   160710.1-391104 & Sz~90   & 16:07:10.1 & -39:11:03 & 102-194459 &  $ -5.4 \pm  7.3$ & $-23.8 \pm  4.0$ & \\  
   160711.6-390348 & Sz~91   & 16:07:11.6 & -39:03:48 & 102-194468 &  $-20.3 \pm  7.5$ & $-18.4 \pm  5.6$ & 3 \\  
   160715.2-400342 & Sz~92, Th~22   & 16:07:15.2 & -40:03:42 & 100-191044 &  $ 16.2 \pm  3.2$ & $-34.1 \pm  3.3$ & \\  
   160727.1-391601 &                & 16:07:27.1 & -39:16:01 & 102-194531 &  $-19.7 \pm  2.3$ & $-24.7 \pm  7.4$ & \\  
   160734.3-392742 &                & 16:07:34.3 & -39:27:42 & 102-194554 &  $ 18.8 \pm  8.5$ & $-22.9 \pm  8.3$ & \\  
   160749.6-390429 & Sz~94   & 16:07:49.6 & -39:04:29 & 102-194608 &  $  0.4 \pm  4.1$ & $-26.7 \pm  4.0$ & \\  
   160754.1-392046 &                & 16:07:54.1 & -39:20:46 & 102-194634 &  $  6.3 \pm 15.6$ & $-19.8 \pm 22.0$ & 1 \\ 
   160758.7-392109 &                & 16:07:58.7 & -39:21:09 & 102-194657 &  $  7.9 \pm 11.3$ & $  3.4 \pm  9.4$ & \\  
   160812.6-390834 & Sz~96    & 16:08:12.6 & -39:08:33 & 102-194723 &  $ -9.1 \pm  3.1$ & $-22.2 \pm  2.8$ & \\
   160817.4-390105 &                & 16:08:17.4 & -39:01:05 & 102-194744 &  $-18.4 \pm  6.9$ & $-29.1 \pm 12.7$ & \\  
   160821.8-390422 & Sz~97, Th~24   & 16:08:21.8 & -39:04:21 & 102-194777 &  $-11.9 \pm  4.6$ & $-22.7 \pm  3.7$ & \\  
   160822.5-390446 & Sz~98, HK~Lup   & 16:08:22.5 & -39:04:46 & 102-194779 &  $ -8.7 \pm  3.6$ & $-23.8 \pm  3.4$ & \\  
   160822.8-390058 &                & 16:08:22.8 & -39:00:58 & 102-194781 &  $-11.6 \pm  3.6$ & $-18.8 \pm  3.5$ & \\
   160825.2-384055 &                & 16:08:25.2 & -38:40:55 & 103-203004 &  $-13.7 \pm  3.5$ & $-24.1 \pm  5.6$ & \\  
   160825.8-390601 & Sz~100, Th~26 & 16:08:25.8 & -39:06:01 & 102-194798 &  $-19.6 \pm 19.1$ & $-21.9 \pm  5.1$ & 2 \\
   160827.8-390040 &                & 16:08:27.8 & -39:00:40 & 102-194810 &  $ -9.1 \pm  5.8$ & $-19.9 \pm  3.8$ & \\  
   160828.4-390532 & Sz~101, Th~27 & 16:08:28.4 & -39:05:32 & 102-194814 &  $-13.7 \pm 10.1$ & $-13.7 \pm  9.8$ & 1 \\
   160829.7-390311 & Sz~102, Krautter's star  & 16:08:29.7 & -39:03:11 & 102-194820 &  $  4.6 \pm  9.3$ & $-31.0 \pm  9.3$ & \\  
   160830.3-390611 & Sz~103, Th~29  & 16:08:30.3 & -39:06:11 & 102-194825 &  $-15.3 \pm 16.1$ & $-13.7 \pm  8.6$ & 1 \\
   160830.7-382827 &                 & 16:08:30.7 & -38:28:27 & 104-202988 &  $ -7.0 \pm  2.1$ & $-25.0 \pm  4.3$ & \\  
   160831.6-384729 &                 & 16:08:31.6 & -38:47:29 & 103-203065 &  $-12.3 \pm  6.7$ & $-22.4 \pm  3.0$ & \\ 
   160836.2-392302 &                 & 16:08:36.2 & -39:23:02 & 102-194850 &  $ -9.2 \pm  3.6$ & $-25.7 \pm  3.2$ & \\  
   160839.8-390625 & Sz~106  & 16:08:39.8 & -39:06:25 & 102-194868 &  $-16.5 \pm 16.1$ & $-38.8 \pm 17.4$ & \\  
   160839.8-392922 &                 & 16:08:39.8 & -39:29:22 & 102-194869 &  $-19.8 \pm  7.8$ & $-32.1 \pm  6.9$ & \\  
   160841.8-390137 & Sz~107  & 16:08:41.8 & -39:01:37 & 102-194875 &  $-11.3 \pm  6.3$ & $-24.4 \pm  6.2$ & \\  
   160851.6-390318 & Sz~110, Th~32  & 16:08:51.6 & -39:03:18 & 102-194906 &  $ -8.7 \pm  3.7$ & $-24.6 \pm  3.7$ & \\  
   160853.2-391440 & 2MASS~J16085324-3914401 & 16:08:53.2 & -39:14:40 & 102-194910 &  $-10.8 \pm 11.4$ & $-27.9 \pm 11.3$ & \\  
   160854.7-393744 & Sz~111, Th~33  & 16:08:54.7 & -39:37:44 & 101-192059 &  $-10.6 \pm  3.1$ & $-21.6 \pm  3.1$ & \\  
   160855.5-390234 & Sz~112  & 16:08:55.5 & -39:02:34 & 102-194917 &  $-11.8 \pm  9.2$ & $-18.0 \pm  5.7$ & 2 \\  
   160901.8-390513 & Sz~114, V908~Sco  & 16:09:01.8 & -39:05:12 & 102-194945 &  $-13.1 \pm  4.9$ & $-22.0 \pm  3.5$ & \\  
   160904.6-392112 &                 & 16:09:04.6 & -39:21:12 & 102-194958 &  $-13.7 \pm  3.8$ & $-20.8 \pm  4.1$ & \\  
   160906.2-390852 & Sz~115  & 16:09:06.2 & -39:08:52 & 102-194964 &  $-16.8 \pm  4.7$ & $-23.4 \pm  4.9$ & \\  
   160915.7-385139 &                 & 16:09:15.7 & -38:51:39 & 103-203455 &  $-21.6 \pm  7.2$ & $-24.2 \pm  4.4$ &  \\
   160937.4-391044 &                 & 16:09:37.4 & -39:10:44 & 102-195079 &  $ -3.4 \pm  3.7$ & $ -1.8 \pm  3.3$ & \\  
   160942.6-391941 & Sz~116, Th~36  & 16:09:42.6 & -39:19:41 & 102-195093 &  $-23.8 \pm  3.1$ & $-35.1 \pm  3.6$ & 2 \\  
   160944.3-391330 & Sz~117, Th~37  & 16:09:44.3 & -39:13:30 & 102-195097 &  $-14.7 \pm  3.5$ & $-26.6 \pm  3.5$ & \\  
   160948.6-391117 & Sz~118  & 16:09:48.6 & -39:11:17 & 102-195106 &  $-19.8 \pm  7.5$ & $-16.9 \pm  7.5$ & \\ 
   160957.1-385948 & Sz~119  & 16:09:57.1 & -38:59:48 & 103-203753 &  $-15.4 \pm  3.3$ & $-34.0 \pm 14.9$ & \\  
   161012.2-392118 & Sz~121, Th~40  & 16:10:12.2 & -39:21:18 & 102-195162 &  $ -9.3 \pm  3.5$ & $-23.8 \pm  3.5$ & \\  
   161016.4-390805 & Sz~122, Th~41   & 16:10:16.4 & -39:08:05 & 102-195173 &  $-17.6 \pm  3.2$ & $-24.0 \pm  3.0$ & \\  
   161034.3-381031 &                & 16:10:34.3 & -38:10:31 & 104-203985 &  $-22.0 \pm  8.9$ & $-15.7 \pm  2.2$ & \\  
   161041.9-382304 &                & 16:10:41.9 & -38:23:04 & 104-204037 &  $-11.2 \pm  3.5$ & $-35.2 \pm  3.4$ & \\  
   161051.6-385314 & Sz~123, Th~42 & 16:10:51.6 & -38:53:14 & 103-204003 &  $ -4.1 \pm  5.9$ & $-21.2 \pm  3.5$ & 2 \\  
   161118.7-385824 &                & 16:11:18.7 & -38:58:24 & 103-204143 &  $-11.9 \pm  3.9$ & $-10.1 \pm  2.9$ & \\  
   161138.1-384135 &                & 16:11:38.1 & -38:41:35 & 103-204233 &  $-15.9 \pm  2.6$ & $-18.2 \pm  2.4$ & \\  
   161153.4-390216 & Sz~124, Th~43 & 16:11:53.3 & -39:02:16 & 102-195419 &  $-19.0 \pm  2.1$ & $-20.9 \pm  2.1$ & \\
   161207.6-381324 &                & 16:12:07.6 & -38:13:24 & 104-204340 &  $ -9.9 \pm  3.8$ & $-20.2 \pm  2.4$ & \\  
   161243.8-381503 &                & 16:12:43.7 & -38:15:03 & 104-204438 &  $ -8.0 \pm  1.9$ & $-16.5 \pm  1.9$ & 1 \\

  \noalign{\smallskip}\hline\noalign{\smallskip}
  \multicolumn{8}{c}{Lupus~4 Cloud}\\
  \noalign{\smallskip}
                                     & Sz~128 & 15:58:07.4 & -41:51:48 & 097-206859 &  $-13.5 \pm  4.8$ & $-20.5 \pm  4.1$ & \\ 
                                     & Sz~129 & 15:59:16.5 & -41:57:09 & 097-207576 &  $ -8.7 \pm  2.1$ & $-21.0 \pm  2.1$ & \\
   160000.6-422158 &                 & 16:00:00.6 & -42:21:57 & 096-205744 &  $ -9.0 \pm  4.6$ & $-20.1 \pm  4.5$ & 2 \\  
   160002.4-422216 &                 & 16:00:02.4 & -42:22:15 & 096-205752 &  $-12.7 \pm  5.4$ & $-25.6 \pm  3.4$ & 2 \\  
   160031.1-414337 & Sz~130  & 16:00:31.0 & -41:43:37 & 097-208086 &  $ -1.8 \pm 10.7$ & $-18.8 \pm  4.5$ & 1, 2 \\  
   160044.5-415531 & MY~Lup, IRAS~15573-4147 & 16:00:44.5 & -41:55:31 & 097-208155 &  $-11.4 \pm  4.3$ & $-22.9 \pm  2.0$ & \\  
   160049.4-413004 & Sz~131  & 16:00:49.4 & -41:30:04 & 097-208172 &  $ -8.9 \pm  5.2$ & $-31.8 \pm  6.2$ & \\  
   160329.4-414003 & Sz~133  & 16:03:29.4 & -41:40:03 & 097-208899 &  $ -4.0 \pm  7.2$ & $-36.0 \pm  7.2$ & 2 \\  
	                           & Sz~134   & 16:09:12.2 & -41:40:25 & 097-211385 &  $-14.3 \pm  3.0$ & $-19.5 \pm 14.7$ & \\
\end{longtable}
\begin{flushleft}
{\bf Notes.} \\
     (1) Ambiguous source (could belong to either group). \\
     (2) This source has a reported visual companion without UCAC3 counterpart. \\
     (3) Reported visual companion (Ghez et al. \cite{ghez97}, $\sim8.9\arcsec$) with UCAC3 counterpart.
\end{flushleft}
}


\addtocounter{table}{1}  

\longtabL{3}{
\begin{longtable}{cc cc c rr c}

\caption{~~~~~~~~~~~~~~~~~~~~~~~~~~~~~~~~~Table~\ref{tab:groupB}: Group B sources}
\label{tab:groupB}\\
    \hline \hline \noalign{\smallskip}
  \multicolumn{1}{c}{SSTc2d~J} &
   \multicolumn{1}{c}{Other names} &
   \multicolumn{1}{c}{$\alpha$ (J2000)} & 
    \multicolumn{1}{c}{$\delta$ (J2000)}  &
   \multicolumn{1}{c}{UCAC3} &
 \multicolumn{1}{c}{$\mu_{\alpha}\cos\delta$} &
  \multicolumn{1}{c}{$\mu_{\delta}$} &
   \multicolumn{1}{c}{Notes} \\
  \multicolumn{1}{l}{} &
   \multicolumn{1}{c}{} &
   \multicolumn{1}{c}{} &
   \multicolumn{1}{c}{} &
   \multicolumn{1}{c}{} &
 \multicolumn{1}{c}{(mas/yr)} &
  \multicolumn{1}{c}{(mas/yr)} &
   \multicolumn{1}{c}{} \\
      \noalign{\smallskip}\hline\noalign{\smallskip}
\endfirsthead

\caption{~~~~~~~~~~~~~~~~~~~~~~~~~~~~~~~~~Table~\ref{tab:groupB}: continued.}\\
   \hline\hline\noalign{\smallskip}
  \multicolumn{1}{c}{SSTc2d~J} &
   \multicolumn{1}{c}{Other names} &
   \multicolumn{1}{c}{$\alpha$ (J2000)} & 
    \multicolumn{1}{c}{$\delta$ (J2000)}  &
   \multicolumn{1}{c}{UCAC3} &
 \multicolumn{1}{c}{$\mu_{\alpha}\cos\delta$} &
  \multicolumn{1}{c}{$\mu_{\delta}$} &
   \multicolumn{1}{c}{Notes} \\
  \multicolumn{1}{l}{} &
   \multicolumn{1}{c}{} &
   \multicolumn{1}{c}{} &
   \multicolumn{1}{c}{} &
   \multicolumn{1}{c}{} &
 \multicolumn{1}{c}{(mas/yr)} &
  \multicolumn{1}{c}{(mas/yr)} &
   \multicolumn{1}{c}{} \\
    \noalign{\smallskip}\hline\noalign{\smallskip}
\endhead
\hline
\endfoot

  \multicolumn{8}{c}{Lupus~1 Cloud}\\
  \noalign{\smallskip}
  153652.8-340956 &  	   & 15:36:52.8 & -34:09:56 & 112-180962 & $  -5.2 \pm  1.7$ & $ -0.1 \pm  1.7$ & \\
  153654.1-342307 &  	   & 15:36:54.1 & -34:23:07 & 112-180966 & $  -7.1 \pm  1.8$ & $  0.5 \pm  1.8$ & \\
  153659.4-341425 &  	   & 15:36:59.4 & -34:14:25 & 112-180979 & $  -5.1 \pm  1.6$ & $ -2.0 \pm  1.6$ & \\
  153736.0-345902 &  	   & 15:37:36.0 & -34:59:02 & 111-179553 & $   0.6 \pm  1.7$ & $  5.3 \pm  1.7$ & \\
  153742.9-332034 &  	   & 15:37:42.9 & -33:20:34 & 114-181267 & $  -4.1 \pm  1.4$ & $  0.2 \pm  1.4$ & \\
  153745.7-342920 &  	   & 15:37:45.7 & -34:29:20 & 112-181159 & $  -9.0 \pm  1.1$ & $  4.1 \pm  1.1$ & \\
  153750.9-332722 &  	   & 15:37:50.9 & -33:27:22 & 114-181289 & $  -7.0 \pm  1.7$ & $  0.1 \pm  1.7$ & \\
  153753.2-343256 &  	   & 15:37:53.2 & -34:32:56 & 111-179602 & $ -12.6 \pm  1.8$ & $  2.4 \pm  1.8$ & \\
  153756.8-342117 &  	   & 15:37:56.8 & -34:21:17 & 112-181197 & $  -0.6 \pm  1.7$ & $  5.8 \pm  2.1$ & \\
  153805.7-341535 &  	   & 15:38:05.7 & -34:15:35 & 112-181227 & $  -5.6 \pm  1.5$ & $ -3.3 \pm  1.4$ & \\
  153808.7-343557 &  	   & 15:38:08.7 & -34:35:57 & 111-179635 & $  -6.5 \pm  1.7$ & $ -0.6 \pm  1.7$ & \\
  153820.2-350215 &  	   & 15:38:20.2 & -35:02:15 & 110-179447 & $   3.2 \pm  1.6$ & $  0.8 \pm  1.7$ & \\
  153821.4-341519 &  	   & 15:38:21.4 & -34:15:19 & 112-181280 & $  -6.6 \pm  3.5$ & $  1.8 \pm  3.6$ & \\
  153824.8-340629 &  	   & 15:38:24.8 & -34:06:29 & 112-181300 & $  -3.9 \pm  1.6$ & $  1.5 \pm  2.9$ & \\
  153831.6-331031 &  	   & 15:38:31.6 & -33:10:31 & 114-181385 & $  -5.9 \pm  1.8$ & $ -4.1 \pm  1.8$ & \\
  153859.5-343458 &  	   & 15:38:59.5 & -34:34:58 & 111-179723 & $ -22.2 \pm  1.5$ & $ -7.3 \pm  1.5$ & 1 \\
  153900.3-345534 &  	   & 15:39:00.3 & -34:55:34 & 111-179724 & $  -8.1 \pm  2.0$ & $ -4.3 \pm  2.0$ & \\
  153902.1-342040 &  	   & 15:39:02.1 & -34:20:40 & 112-181429 & $  -6.9 \pm  2.3$ & $ -4.4 \pm  2.3$ & \\
  153905.2-341210 &  	   & 15:39:05.2 & -34:12:10 & 112-181437 & $  -8.3 \pm  2.1$ & $ -6.6 \pm  2.3$ & \\
  153922.3-334909 &  	   & 15:39:22.3 & -33:49:09 & 113-182842 & $ -11.0 \pm  2.3$ & $ -0.7 \pm  2.3$ & \\
  153929.0-340339 &  	   & 15:39:29.0 & -34:03:39 & 112-181511 & $  -7.9 \pm  2.3$ & $ -3.8 \pm  2.3$ & \\
  153929.2-334322 &  	   & 15:39:29.2 & -33:43:22 & 113-182854 & $ -16.1 \pm  3.1$ & $ -1.0 \pm  2.3$ & \\
  153930.5-342058 &  	   & 15:39:30.5 & -34:20:58 & 112-181518 & $  -0.6 \pm  2.0$ & $ -2.2 \pm  2.0$ & \\
  153932.6-335205 &  	   & 15:39:32.6 & -33:52:05 & 113-182872 & $  -6.6 \pm  2.3$ & $ -3.5 \pm  2.3$ & \\
  153933.2-350151 &  	   & 15:39:33.2 & -35:01:51 & 110-179659 & $  -6.0 \pm  4.5$ & $  4.5 \pm  2.4$ & \\
  153943.8-341506 &  	   & 15:39:43.8 & -34:15:06 & 112-181558 & $  -3.7 \pm  1.7$ & $  2.4 \pm  1.7$ & \\
  153950.8-340456 &  	   & 15:39:50.8 & -34:04:56 & 112-181575 & $  -8.8 \pm  2.5$ & $ -5.1 \pm  2.5$ & \\
  153955.6-341103 &  	   & 15:39:55.6 & -34:11:03 & 112-181587 & $ -21.5 \pm  2.3$ & $ -2.9 \pm  2.3$ & \\
  154006.1-343224 &  	   & 15:40:06.1 & -34:32:24 & 111-179825 & $  -9.4 \pm  2.7$ & $ -3.5 \pm  2.4$ & \\
  154006.9-332331 &  	   & 15:40:06.9 & -33:23:31 & 114-181587 & $  -3.2 \pm  2.7$ & $ -2.4 \pm  2.4$ & \\
  154013.7-340142 &  	   & 15:40:13.7 & -34:01:42 & 112-181650 & $  -4.7 \pm  2.3$ & $-15.1 \pm  2.1$ & \\
  154033.8-342607 &  	   & 15:40:33.8 & -34:26:07 & 112-181729 & $  -4.5 \pm  1.7$ & $ -9.8 \pm  1.6$ &  \\
  154038.1-341417 &  	   & 15:40:38.1 & -34:14:17 & 112-181753 & $ -14.3 \pm  8.6$ & $  0.0 \pm  2.5$ & \\
  154040.4-340455 &  	   & 15:40:40.4 & -34:04:55 & 112-181764 & $ -13.6 \pm  1.8$ & $ -3.5 \pm  1.4$ & \\
  154047.2-331347 &  	   & 15:40:47.2 & -33:13:47 & 114-181775 & $ -11.3 \pm  3.2$ & $ -6.8 \pm  1.7$ & \\
  154053.7-342745 &  	   & 15:40:53.7 & -34:27:45 & 112-181810 & $  -2.2 \pm  2.8$ & $  2.0 \pm  2.3$ & \\
  154116.9-335544 &  	   & 15:41:16.9 & -33:55:44 & 113-183184 & $   2.4 \pm  2.3$ & $ -7.1 \pm  2.3$ & \\
  154118.0-344301 &  	   & 15:41:18.0 & -34:43:01 & 111-179990 & $ -12.2 \pm  1.9$ & $ -7.7 \pm  1.9$ & 1 \\
  154126.7-334732 &  	   & 15:41:26.7 & -33:47:32 & 113-183205 & $  -2.2 \pm  2.0$ & $ -8.5 \pm  2.0$ & \\
  154130.4-342759 &  	   & 15:41:30.4 & -34:27:59 & 112-181935 & $  -6.8 \pm  2.2$ & $  3.2 \pm  2.2$ & \\
  154136.0-342532 &  	   & 15:41:36.0 & -34:25:32 & 112-181963 & $  -0.7 \pm  2.4$ & $ -0.6 \pm  2.4$ & \\
  154136.0-343818 &  	   & 15:41:36.0 & -34:38:18 & 111-180053 & $ -13.8 \pm  1.9$ & $ -3.0 \pm  1.9$ & \\
  154149.5-334517 &  	   & 15:41:49.5 & -33:45:17 & 113-183262 & $  -5.0 \pm  2.5$ & $  0.5 \pm  2.4$ & \\
  154151.7-335809 &  	   & 15:41:51.7 & -33:58:09 & 113-183268 & $ -21.1 \pm  2.6$ & $ -1.7 \pm  1.7$ & \\
  154201.8-345449 &  	   & 15:42:01.8 & -34:54:49 & 111-180129 & $ -17.9 \pm  2.2$ & $  2.2 \pm  2.2$ & \\
  154344.5-335834 &  	   & 15:43:44.5 & -33:58:34 & 113-183752 & $ -43.9 \pm  1.6$ & $-11.5 \pm  1.6$ & \\ 
  154401.4-340229 &  	   & 15:44:01.4 & -34:02:29 & 112-182317 & $  -5.8 \pm  2.3$ & $  0.9 \pm  2.3$ & \\
  154406.0-343238 &  	   & 15:44:06.0 & -34:32:38 & 111-180557 & $  -7.7 \pm  2.6$ & $ -3.2 \pm  2.5$ & \\
  154624.2-343443 &  	   & 15:46:24.2 & -34:34:43 & 111-180973 & $  -6.5 \pm  4.4$ & $ -6.6 \pm  2.3$ & \\
  154637.4-344307 &  	   & 15:46:37.4 & -34:43:07 & 111-181011 & $ -12.2 \pm  5.7$ & $ -0.4 \pm  3.1$ & \\

  \noalign{\smallskip}\hline\noalign{\smallskip}
  \multicolumn{8}{c}{Lupus~3 Cloud}\\
  \noalign{\smallskip}

  160509.3-391203 &  	                            & 16:05:09.3 & -39:12:03 & 102-193467 & $ -10.4 \pm  7.5$ & $ -8.2 \pm  3.1$ & \\
  160624.4-392158 &  	                            & 16:06:24.4 & -39:21:58 & 102-194148 & $   0.0 \pm  2.8$ & $ -5.9 \pm  2.8$ & \\
  160659.5-390605 &  	                            & 16:06:59.5 & -39:06:05 & 102-194390 & $  -1.4 \pm  2.8$ & $-13.4 \pm  5.7$ & 1 \\
  160702.6-391203 &  	                            & 16:07:02.6 & -39:12:03 & 102-194415 & $  -6.4 \pm  3.2$ & $  2.6 \pm  3.2$ & \\
  160708.6-394723 &  	                            & 16:07:08.6 & -39:47:23 & 101-190770 & $ -14.0 \pm  5.4$ & $ -2.8 \pm  3.3$ & 2 \\
  160713.5-383525 &  	                            & 16:07:13.5 & -38:35:25 & 103-202192 & $  -1.9 \pm 17.8$ & $  0.4 \pm  4.6$ & \\
                                    & Sz~93                       & 16:07:17.8 & -39:34:05 & 101-190887 & $  -3.0 \pm  2.7$ & $ -9.0 \pm  2.7$ & \\
  160732.3-383508 &  	                            & 16:07:32.3 & -38:35:08 & 103-202425 & $  -7.4 \pm  3.3$ & $ -0.8 \pm  3.3$ & \\
  160734.3-392742 &  	                            & 16:07:34.3 & -39:27:42 & 102-194554 & $  18.8 \pm  8.5$ & $-22.9 \pm  8.3$ & \\
  160735.3-392507 &  	                            & 16:07:35.3 & -39:25:07 & 102-194557 & $  -0.9 \pm  3.6$ & $  6.1 \pm  5.1$ & \\
  160745.9-385245 &  	                            & 16:07:45.9 & -38:52:45 & 103-202567 & $ -14.2 \pm  7.2$ & $ -1.0 \pm  6.6$ & \\
  160747.4-392606 &  	                            & 16:07:47.4 & -39:26:06 & 102-194597 & $  -5.1 \pm  2.7$ & $ -4.7 \pm  4.0$ & \\
  160758.7-392109 &  	                            & 16:07:58.7 & -39:21:09 & 102-194657 & $   7.9 \pm 11.3$ & $  3.4 \pm  9.4$ & \\
  160803.0-385229 &  	                            & 16:08:03.0 & -38:52:29 & 103-202756 & $  17.4 \pm  3.3$ & $ -3.8 \pm  3.3$ & 2 \\
  160804.6-384558 &  	                            & 16:08:04.6 & -38:45:58 & 103-202775 & $   3.9 \pm  2.4$ & $ -0.3 \pm  2.4$ & \\
  160829.3-383551 &  	                            & 16:08:29.3 & -38:35:51 & 103-203038 & $  -3.5 \pm  8.6$ & $ -4.8 \pm  7.8$ & \\
  160835.8-390348 & Par-Lup3-2 	         & 16:08:35.8 & -39:03:48 & 102-194847 & $  -7.8 \pm  5.7$ & $-10.6 \pm 7.5$ & 1 \\
                                    & Sz~105                     & 16:08:37.0 & -40:16:21 & 100-192100 & $   0.2 \pm  6.9$ & $ -8.6 \pm  3.7$ & 3 \\
  160844.3-374443 &  	                            & 16:08:44.3 & -37:44:43 & 105-204594 & $   3.1 \pm  1.4$ & $ -1.8 \pm  1.4$ & \\
  160846.4-393347 &  	                            & 16:08:46.4 & -39:33:47 & 101-191965 & $   0.5 \pm 15.6$ & $ -7.5 \pm  6.2$ & \\
  160846.6-384112 &  	                            & 16:08:46.6 & -38:41:12 & 103-203212 & $   6.0 \pm  9.2$ & $ -1.4 \pm  3.6$ & \\
  160855.6-392316 &  	                            & 16:08:55.6 & -39:23:16 & 102-194918 & $   1.0 \pm 11.4$ & $ -6.9 \pm  4.1$ & \\
  160857.3-392849 &  	                            & 16:08:57.3 & -39:28:49 & 102-194927 & $  -3.9 \pm  3.1$ & $  0.7 \pm  3.1$ & \\
  160903.7-385610 &  	                            & 16:09:03.7 & -38:56:10 & 103-203354 & $  -0.9 \pm  3.4$ & $  2.0 \pm  2.8$ & \\
  160903.8-384126 &  	                            & 16:09:03.8 & -38:41:26 & 103-203355 & $  -4.6 \pm  3.5$ & $  1.7 \pm  2.6$ & \\
  160917.6-392537 &  	                            & 16:09:17.6 & -39:25:37 & 102-195004 & $   0.8 \pm  3.0$ & $ -2.5 \pm  2.8$ & \\
  160934.1-391342 &  	                            & 16:09:34.1 & -39:13:42 & 102-195072 & $  -1.1 \pm  5.6$ & $ -9.0 \pm  5.6$ & 1 \\
  160937.4-391044 &  	                            & 16:09:37.4 & -39:10:44 & 102-195079 & $  -3.4 \pm  3.7$ & $ -1.8 \pm  3.3$ & \\
  160939.5-384431 &  	                            & 16:09:39.5 & -38:44:31 & 103-203637 & $   5.4 \pm  7.9$ & $-11.7 \pm  6.1$ & 1 \\
  161013.5-384208 &  	                            & 16:10:13.5 & -38:42:08 & 103-203833 & $  -1.6 \pm  3.2$ & $ -9.0 \pm  2.7$ & 1 \\
  161032.6-374615 & IRAS~16072-3738 & 16:10:32.6 & -37:46:15 & 105-205237 & $  -7.7 \pm  6.8$ & $  5.4 \pm  3.2$ & \\
  161033.2-383023 &  	                            & 16:10:33.2 & -38:30:23 & 103-203926 & $  -3.1 \pm  6.5$ & $-12.0 \pm  8.3$ & 1 \\
  161034.5-381450 &  	                            & 16:10:34.5 & -38:14:50 & 104-203987 & $   6.1 \pm  3.0$ & $ -3.0 \pm  7.6$ & \\
  161036.7-380927 &  	                            & 16:10:36.7 & -38:09:27 & 104-204003 & $  -9.0 \pm  1.4$ & $-12.0 \pm  2.4$ & 1 \\
  161114.8-384618 &  	                            & 16:11:14.8 & -38:46:18 & 103-204118 & $  -5.4 \pm  6.1$ & $  1.6 \pm  2.5$ & \\
  161118.7-385824 &  	                            & 16:11:18.7 & -38:58:24 & 103-204143 & $ -11.9 \pm  3.9$ & $-10.1 \pm  2.9$ & 1, 2 \\
  161148.7-381758 &  	                            & 16:11:48.7 & -38:17:58 & 104-204298 & $  -1.2 \pm  6.1$ & $ -6.8 \pm 15.5$ & \\
  161200.1-385557 &  	                            & 16:12:00.1 & -38:55:57 & 103-204299 & $   8.0 \pm 18.5$ & $ -7.2 \pm  4.8$ & 2 \\
  161200.9-383625 &  	                            & 16:12:00.9 & -38:36:25 & 103-204302 & $  -0.1 \pm  4.2$ & $  6.1 \pm  2.3$ & \\
  161219.6-383742 &  	                            & 16:12:19.6 & -38:37:42 & 103-204359 & $  -0.4 \pm  8.6$ & $ -5.7 \pm  3.3$ & \\
                                    & Sz~125                     & 16:12:30.1 & -39:35:40 & 101-193738 & $  -8.9 \pm  3.5$ & $  1.7 \pm  2.4$ & \\
  161251.7-384216 &  	                            & 16:12:51.7 & -38:42:16 & 103-204499 & $   0.6 \pm  3.1$ & $-11.5 \pm  8.1$ & 1 \\
  161256.0-375643 &  	                            & 16:12:56.0 & -37:56:44 & 105-205869 & $  -8.8 \pm 10.4$ & $  0.1 \pm  4.0$ & \\
  161341.0-383724 &  	                            & 16:13:40.9 & -38:37:24 & 103-204732 & $ -15.3 \pm 10.4$ & $  8.7 \pm  7.5$ & \\

 \noalign{\smallskip}\hline\noalign{\smallskip}
  \multicolumn{8}{c}{Lupus~4 Cloud}\\
  \noalign{\smallskip}
  155921.8-412808 &  	                            & 15:59:21.8 & -41:28:08 & 098-201434 & $ -12.0 \pm  5.6$ & $ -8.0 \pm  2.8$ & \\
  160143.3-413606 &  	                            & 16:01:43.3 & -41:36:06 & 097-208357 & $  -5.9 \pm 12.0$ & $ -4.0 \pm  5.6$ & \\
  160157.0-414244 & IRAS~15585-4134 & 16:01:57.0 & -41:42:44 & 097-208409 & $  -5.8 \pm  4.7$ & $ -8.5 \pm  4.6$ & 1 \\
\end{longtable}
\begin{flushleft}
{\bf Notes.} \\
     (1) Ambiguous source (could belong to either group). \\
     (2) This source has a reported visual companion without UCAC3 counterpart. \\
     (3) Reported visual companion (Ghez et al. \cite{ghez97}, $\sim10\arcsec$) with UCAC3 counterpart.
\end{flushleft}
}

\addtocounter{table}{1}  
\longtabL{4}{
\begin{longtable}{cc cc ccc cc}

\caption{~~~~~~~~~~~~~~~~~~~~~~~~~~~~~~~~~~~Table~\ref{tab:groupAphot}: Photometry for Group A sources (Lupus kinematic members)}
\label{tab:groupAphot}\\
    \hline \hline \noalign{\smallskip}
  \multicolumn{1}{c}{SSTc2d~J} &
   \multicolumn{1}{c}{Other names} &
  \multicolumn{1}{c}{Rc} &
  \multicolumn{1}{c}{Ic} &
  \multicolumn{1}{c}{J} &
  \multicolumn{1}{c}{H} &
  \multicolumn{1}{c}{Ks} &
  \multicolumn{1}{c}{Class} &
  \multicolumn{1}{c}{SpT}  \\
  \multicolumn{1}{l}{} &
  \multicolumn{1}{l}{} &
  \multicolumn{1}{c}{(mag)} &
  \multicolumn{1}{c}{(mag)} &
  \multicolumn{1}{c}{(mag)} &
  \multicolumn{1}{c}{(mag)} &
  \multicolumn{1}{c}{(mag)} &
  \multicolumn{1}{c}{} &
  \multicolumn{1}{c}{} \\
      \noalign{\smallskip}\hline\noalign{\smallskip}
\endfirsthead

\caption{~~~~~~~~~~~~~~~~~~~~~~~~~~~~~~~~~~~Table~\ref{tab:groupAphot}: continued.}\\
   \hline\hline\noalign{\smallskip}
  \multicolumn{1}{c}{SSTc2d~J} &
   \multicolumn{1}{c}{Other names} &
  \multicolumn{1}{c}{Rc} &
  \multicolumn{1}{c}{Ic} &
  \multicolumn{1}{c}{J} &
  \multicolumn{1}{c}{H} &
  \multicolumn{1}{c}{Ks} &
  \multicolumn{1}{c}{Class} &
  \multicolumn{1}{c}{SpT} \\
  \multicolumn{1}{l}{} &
   \multicolumn{1}{c}{} &
  \multicolumn{1}{c}{(mag)} &
  \multicolumn{1}{c}{(mag)} &
  \multicolumn{1}{c}{(mag)} &
  \multicolumn{1}{c}{(mag)} &
  \multicolumn{1}{c}{(mag)} &
  \multicolumn{1}{c}{} &
  \multicolumn{1}{c}{}  \\
    \noalign{\smallskip}\hline\noalign{\smallskip}
\endhead
\hline
\endfoot

  \multicolumn{9}{c}{Lupus~1 Cloud}\\
  \noalign{\smallskip}
  153805.7-341535 &  	           	     & 13.50 & 11.66 &        &        &	& III	 &	 \\
  153905.2-341210 &  	           	     & 13.47 & 11.69 &        &        &	& III	 &	 \\
  153927.8-344617 & Sz~65, IK~Lup  	     & 11.33 & 10.76 &  9.189 &  8.414 & 7.982  & II	 & M0	 \\
  154009.4-342734 &                	     & 13.48 & 11.68 & 10.044 &  9.56  & 9.323  & III	 &	 \\
  154013.7-340142 &                	     & 13.81 & 12.21 & 10.734 & 10.052 & 9.852  & III	 &	 \\			    
  154018.5-342614 &                	     & 14.44 & 12.61 & 11.015 & 10.43  & 10.166 & III	 &	 \\
  154122.0-344015 &                	     & 14.23 & 12.61 & 11.084 & 10.414 & 10.21  & III	 &	 \\
  154148.3-350145 &                	     & 14.34 & 12.6  & 11.049 & 10.446 & 10.228 & III	 &	 \\
  154201.8-345449 &                	     & 13.48 & 11.96 &        &        &	& III	 &	 \\			       
  154257.6-341928 &  	           	     & 15.02 & 13.21 & 11.233 & 10.411 & 10.094 & III	 &	 \\
  154344.5-335834 &  	           	     & 13.37 & 11.62 &        &        &	& III	 &	 \\
  154512.9-341731 & Sz~68, HT~Lup  	     &       &       &  7.573 &  6.866 & 6.48	& II	 & K2	 \\
  154517.4-341829 & Sz~69, HW~Lup  	     &       &       & 11.176 & 10.162 & 9.41	& II	 & M1	 \\
  	          & Sz~70          	     &       &       & 10.845 & 10.171 & 9.8	& III	 & M5	 \\
  	          & Sz~71          	     &       &       & 10.073 &  9.184 & 8.63	& II	 & M2	 \\
  	          & Sz~72          	     &       &       & 10.574 &  9.766 & 9.325  & II	 & M3	 \\
  	          & Sz~73          	     &       &       & 10.739 &  9.526 & 8.826  & II	 & M0	 \\
  	          & Sz~74          	     &       &       &  9.225 &  8.098 & 7.432  & II	 & M1.5  \\
  	          & Sz~75          	     &       &       &  8.605 &  7.702 & 7.096  & II	 & K7-M0 \\
  	          & Sz~76          	     &       &       & 10.96  & 10.28  & 10.022 & III	 & M1	 \\
  	          & Sz~77          	     &       &       &  9.444 &  8.59  & 8.271  & II	 & M0	 \\

  \noalign{\smallskip}\hline\noalign{\smallskip}
  \multicolumn{9}{c}{Lupus~3 Cloud}\\
  \noalign{\smallskip}
                  & RY~Lup         	     &       &       &  8.546 &  7.69  &  6.976 & II	 & K4	 \\
		  & EX~Lup         	     &       &       &  9.728 & 8.958  &  8.496 & II	 & M0	 \\
                  & Sz~86          	     &       &       & 12.036 & 11.333 & 11.145 &	 &	 \\
  160644.6-390432 &		   	     & 15.19 & 13.1  & 11.087 & 10.478 & 10.164 & III	 &	 \\
  160710.1-391104 & Sz~90          	     & 13.7  & 12.49 & 10.356 &  9.32  &  8.724 & II	 & K7-M0 \\
  160711.6-390348 & Sz~91          	     & 14.11 & 12.92 & 11.055 & 10.124 &  9.848 & III	 & M0.5  \\
  160715.2-400342 & Sz~92, Th~22   	     &       &       & 11.266 & 10.567 & 10.411 &	 &	 \\
  160727.1-391601 &		   	     & 14.02 & 12.07 & 10.516 & 10.003 &  9.797 & III	 &	 \\  
  160734.3-392742 &		   	     & 13.85 & 11.84 &  9.844 &  9.074 &  8.806 & III	 &	 \\
  160749.6-390429 & Sz~94          	     & 14.76 & 13.0  & 11.448 & 10.805 & 10.557 & III	 & M4	 \\
  160754.1-392046 &		   	     & 14.58 & 12.26 &  9.45  &  8.411 &  7.906 & III	 &	 \\
  160758.7-392109 &		   	     & 14.02 & 12.08 &        &        &	& III	 &	 \\		
  160812.6-390834 & Sz~96          	     & 13.69 & 11.83 & 10.128 &  9.348 &  8.957 & III	 & M1.5  \\
  160817.4-390105 &		   	     & 15.21 & 13.31 & 11.471 & 10.838 & 10.535 & III	 &	 \\
  160821.8-390422 & Sz~97, Th~24   	     & 14.67 & 12.92 & 11.242 & 10.551 & 10.216 & II	 & M3	 \\
  160822.5-390446 & Sz~98, HK~Lup  	     & 11.7  & 11.21 &  9.53  &  8.653 &  8.014 & II	 & M0	 \\
  160822.8-390058 &		   	     & 14.48 & 12.75 &        &        &	& II	 &	 \\
  160825.2-384055 &		   	     & 14.75 & 13.06 & 11.408 & 10.725 & 10.498 & III	 &	 \\
  160825.8-390601 & Sz~100, Th~26  	     & 14.97 & 13.12 & 10.982 & 10.353 &  9.908 & II	 & M5	 \\
  160827.8-390040 &	           	     & 14.22 & 12.44 & 10.799 & 10.057 &  9.803 & III	 &	 \\
  160828.4-390532 & Sz~101, Th~27  	     & 14.0  & 12.35 & 10.4   &  9.722 &  9.388 & III	 & M4	 \\
  160829.7-390311 & Sz~102, Krautter's star  & 15.38 & 15.69 & 14.558 & 13.653 & 12.58  & I	 & K0:   \\
  160830.3-390611 & Sz~103, Th~29            & 15.18 & 13.15 & 11.383 & 10.622 & 10.23  & II	 & M4	 \\
  160830.7-382827 &		             & 10.67 &       & 8.974  &  8.389 &  8.225 & III	 &	 \\
  160831.6-384729 &		             & 13.68 & 11.64 & 9.676  &  8.926 &  8.623 & TO/III &	 \\
  160836.2-392302 &		             & 13.83 & 11.78 &  9.884 &  9.043 &  8.658 & I/II   &	 \\
  160839.8-390625 & Sz~106                   & 15.0  & 14.66 & 11.65  & 10.66  & 10.149 & II	 & M2.5  \\
  160839.8-392922 &		             & 15.44 & 13.49 & 11.735 & 11.154 & 10.926 & III	 &	 \\
  160841.8-390137 & Sz~107                   & 15.42 & 13.2  & 11.249 & 10.624 & 10.311 & III	 & M5.5  \\
  160851.6-390318 & Sz~110, Th~32            & 13.72 & 12.28 & 10.967 & 10.22  &  9.746 & II	 & M4.5  \\
  160853.2-391440 & 2MASS~J16085324-3914401  & 15.35 & 13.52 & 11.325 & 10.283 &  9.8	& II	 &	 \\
  160854.7-393744 & Sz~111, Th~33            & 13.29 &       & 10.62  &  9.801 &  9.539 & III	 & M1.5  \\
  160855.5-390234 & Sz~112                   & 14.78 & 12.93 & 11.004 & 10.286 &  9.962 & II	 & M6	 \\
  160901.8-390513 & Sz~114, V908~Sco         & 14.35 & 12.54 & 10.414 &  9.697 &  9.319 & II	 & M5.5  \\
  160904.6-392112 &		             & 14.61 & 12.52 & 10.681 & 10.014 &  9.762 & III	 &	 \\
  160906.2-390852 & Sz~115                   & 15.09 & 13.12 & 11.334 & 10.649 & 10.447 & II	 & M4	 \\
  160915.7-385139 &		             & 15.42 & 13.72 & 12.046 & 11.503 & 11.224 & III	 &	 \\
  160937.4-391044 &		             & 14.42 & 12.73 &        &        &	& III	 &	 \\		
  160942.6-391941 & Sz~116, Th~36            & 13.05 & 12.2  & 10.471 &  9.77  &  9.526 & III	 & M1.5  \\
  160944.3-391330 & Sz~117, Th~37            & 14.2  & 12.38 & 10.675 &  9.843 &  9.423 & II	 & M2	 \\
  160948.6-391117 & Sz~118                   & 16.61 & 15.0  & 10.456 &  9.351 &  8.685 & II	 & K6	 \\
  160957.1-385948 & Sz~119                   & 13.49 & 12.1  & 10.396 &  9.67  &  9.421 & III	 & M4	 \\
  161012.2-392118 & Sz~121, Th~40            & 13.88 & 11.9  & 10.075 &  9.309 &  9.028 & III	 & M3	 \\
  161016.4-390805 & Sz~122, Th~41            & 13.28 & 12.18 & 10.869 & 10.121 &  9.932 & III	 & M2	 \\
  161034.3-381031 &		             & 13.6  & 11.95 & 10.391 &  9.675 &  9.421 & III	 &	 \\
  161041.9-382304 &		             & 14.32 & 12.56 & 10.933 & 10.285 & 10.045 & III	 &	 \\
  161051.6-385314 & Sz~123, Th~42            & 14.44 & 12.98 & 11.092 & 10.207 &  9.78  & II	 & M3	 \\
  161118.7-385824 &		             & 13.97 &       & 10.287 &  9.629 &  9.438 & III	 &	 \\
  161138.1-384135 &		             & 14.17 & 12.42 & 10.912 & 10.233 & 10.055 & III	 &	 \\
  161153.4-390216 & Sz~124, Th~43            & 12.33 & 11.41 & 10.929 & 10.269 &  9.995 & III	 & K7-M0 \\
  161207.6-381324 &		             & 14.42 & 12.72 & 10.544 &  9.762 &  9.539 & III	 &	 \\
  161243.8-381503 &		             & 13.39 & 11.61 &        &        &	& II	 &	 \\

 \noalign{\smallskip}\hline\noalign{\smallskip}
 \multicolumn{9}{c}{Lupus~4 Cloud}\\
 \noalign{\smallskip}
		  & Sz~128                   &       &       & 10.528 &  9.642 & 9.164  & III	& M1.5  \\
		  & Sz~129                   &       &       &  9.933 &  9.083 & 8.608  & II	& K7-M0 \\
  160000.6-422158 &		             & 15.25 & 13.43 & 11.631 & 10.983 & 10.656 & III	&	\\
  160002.4-422216 &		             & 14.98 & 13.59 & 11.449 & 10.59  & 10.142 & III	&	\\
  160031.1-414337 & Sz~130                   & 13.56 & 12.46 & 10.73  &  9.936 & 9.617  & II	& M1.5  \\  
  160044.5-415531 & MY~Lup, IRAS~15573-4147  & 11.06 &       &  9.457 &  8.685 & 8.348  &	&	\\
  160049.4-413004 & Sz~131                   & 14.6  & 12.92 & 11.466 & 10.612 & 10.1   & II	& M2	\\
  160329.4-414003 & Sz~133                   & 15.78 & 15.04 & 12.105 & 10.548 & 9.53   & II	& K2	\\
	          & Sz~134                   &       &       & 10.788 & 10.072 & 9.747  & III	& M1	\\
\end{longtable}
\begin{flushleft}
{\bf Notes.} \\
      References: Hughes et al. (\cite{hughes94}), Mer\'{\i}n et al. (\cite{merin08}), Comer\'on (\cite{comeron08}), Comer\'on et al. (\cite{comeron09}) \\
\end{flushleft}
}

\addtocounter{table}{1}  
\longtabL{5}{
\begin{longtable}{cc cc ccc cc}

\caption{~~~~~~~~~~~~~~~~~~~~~~~~~~~~~~~~~~~~~~~~~~~~~~~~~Table~\ref{tab:groupBphot}: Photometry for Group B sources}
\label{tab:groupBphot}\\
    \hline \hline \noalign{\smallskip}
  \multicolumn{1}{c}{SSTc2d~J} &
   \multicolumn{1}{c}{Other names} &
  \multicolumn{1}{c}{Rc} &
  \multicolumn{1}{c}{Ic} &
  \multicolumn{1}{c}{J} &
  \multicolumn{1}{c}{H} &
  \multicolumn{1}{c}{Ks} &
  \multicolumn{1}{c}{Class} &
  \multicolumn{1}{c}{SpT}  \\
  \multicolumn{1}{l}{} &
  \multicolumn{1}{l}{} &
  \multicolumn{1}{c}{(mag)} &
  \multicolumn{1}{c}{(mag)} &
  \multicolumn{1}{c}{(mag)} &
  \multicolumn{1}{c}{(mag)} &
  \multicolumn{1}{c}{(mag)} &
  \multicolumn{1}{c}{} &
  \multicolumn{1}{c}{} \\
      \noalign{\smallskip}\hline\noalign{\smallskip}
\endfirsthead

\caption{~~~~~~~~~~~~~~~~~~~~~~~~~~~~~~~~~~~~~~~~~~~~~~~~~Table~\ref{tab:groupBphot}: continued.}\\
   \hline\hline\noalign{\smallskip}
  \multicolumn{1}{c}{SSTc2d~J} &
   \multicolumn{1}{c}{Other names} &
  \multicolumn{1}{c}{Rc} &
  \multicolumn{1}{c}{Ic} &
  \multicolumn{1}{c}{J} &
  \multicolumn{1}{c}{H} &
  \multicolumn{1}{c}{Ks} &
  \multicolumn{1}{c}{Class} &
  \multicolumn{1}{c}{SpT} \\
  \multicolumn{1}{l}{} &
   \multicolumn{1}{c}{} &
  \multicolumn{1}{c}{(mag)} &
  \multicolumn{1}{c}{(mag)} &
  \multicolumn{1}{c}{(mag)} &
  \multicolumn{1}{c}{(mag)} &
  \multicolumn{1}{c}{(mag)} &
  \multicolumn{1}{c}{} &
  \multicolumn{1}{c}{}  \\
    \noalign{\smallskip}\hline\noalign{\smallskip}
\endhead
\hline
\endfoot

  \multicolumn{9}{c}{Lupus~1 Cloud}\\
  \noalign{\smallskip}
  153652.8-340956 &  	    	    & 14.0  & 12.08 & 10.395 &  9.506 &  9.196 & III &    \\ 
  153654.1-342307 &  	    	    & 13.88 & 12.45 & 10.997 & 10.172 &  9.942 & III &    \\ 
  153659.4-341425 &  	    	    & 13.53 & 11.83 & 10.317 &  9.531 &  9.262 & III &    \\ 
  153736.0-345902 &  	    	    & 13.9  & 12.29 & 10.679 &  9.865 &  9.637 & III &    \\ 
  153742.9-332034 &  	    	    & 13.65 & 11.77 &  9.952 &  9.184 &  8.972 & III &    \\ 
  153745.7-342920 &  	    	    & 13.64 & 11.56 &  9.749 &  9.07  &  8.853 & III &    \\ 
  153750.9-332722 &  	    	    & 13.76 & 12.03 & 10.368 &  9.575 &  9.321 & III &    \\ 
  153753.2-343256 &  	    	    & 13.87 & 12.35 & 10.82  & 10.037 &  9.813 & III &    \\ 
  153756.8-342117 &  	    	    & 13.78 & 12.23 & 10.795 & 10.037 &  9.84  & III &    \\ 
  153805.7-341535 &  	    	    & 13.5  & 11.66 & 10.036 &  9.268 &  9.083 & III &    \\ 
  153808.7-343557 &  	    	    & 13.62 & 11.88 & 10.286 &  9.451 &  9.233 & III &    \\ 
  153820.2-350215 &  	    	    & 13.7  & 12.05 & 10.416 &  9.602 &  9.34  & III &    \\ 
  153821.4-341519 &  	    	    & 13.91 & 12.4  & 10.955 & 10.187 &  9.96  & III &    \\ 
  153824.8-340629 &  	    	    & 13.64 & 11.52 &  9.5   &  8.647 &  8.437 & III &    \\ 
  153831.6-331031 &  	    	    & 13.9  & 12.19 & 10.436 &  9.566 &  9.296 & III &    \\ 
  153859.5-343458 &  	    	    & 13.44 & 11.54 & 10.004 &  9.485 &  9.248 & III &    \\ 
  153900.3-345534 &  	    	    & 13.4  & 11.76 & 10.201 &  9.474 &  9.233 & III &    \\ 
  153902.1-342040 &  	    	    & 13.76 & 12.26 & 10.746 &  9.904 &  9.712 & III &    \\ 
  153905.2-341210 &  	    	    & 13.47 & 11.69 &  9.961 &  9.163 &  8.9   & III &    \\ 
  153922.3-334909 &  	    	    & 13.89 & 12.27 & 10.82  &  9.974 &  9.76  & III &    \\ 
  153929.0-340339 &  	    	    & 13.89 & 12.36 & 10.884 & 10.128 &  9.898 & III &    \\ 
  153929.2-334322 &  	    	    & 13.82 & 12.06 & 10.408 &  9.562 &  9.327 & III &    \\ 
  153930.5-342058 &  	    	    & 13.66 & 12.02 & 10.494 &  9.807 &  9.609 & III &    \\ 
  153932.6-335205 &  	    	    & 13.98 & 12.24 & 10.492 &  9.552 &  9.319 & III &    \\ 
  153933.2-350151 &  	    	    & 13.9  & 12.29 & 10.588 &  9.708 &  9.458 & III &    \\ 
  153943.8-341506 &  	    	    & 13.61 & 11.96 & 10.495 &  9.856 &  9.694 & III &    \\ 
  153950.8-340456 &  	    	    & 14.13 & 12.57 & 11.026 & 10.162 &  9.937 & III &    \\ 
  153955.6-341103 &  	    	    & 13.57 & 11.83 & 10.173 &  9.35  &  9.148 & III &    \\ 
  154006.1-343224 &  	    	    & 13.78 & 12.2  & 10.626 &  9.923 &  9.671 & III &    \\ 
  154006.9-332331 &  	    	    & 13.45 & 11.75 & 10.295 &  9.422 &  9.157 & III &    \\ 
  154033.8-342607 &  	    	    & 13.47 & 11.79 & 10.273 &  9.653 &  9.471 & III &    \\ 
  154038.1-341417 &  	    	    & 13.5  & 11.97 & 10.461 &  9.666 &  9.448 & III &    \\ 
  154040.4-340455 &  	    	    & 13.56 & 11.54 &  9.81  &  9.184 &  8.994 & III &    \\ 
  154047.2-331347 &  	    	    & 13.99 & 12.5  & 11.061 & 10.223 &  9.953 & III &    \\ 
  154053.7-342745 &  	    	    & 13.74 & 12.24 & 10.769 & 10.033 &  9.872 & III &    \\ 
  154116.9-335544 &  	    	    & 13.71 & 12.07 & 10.508 &  9.776 &  9.573 & III &    \\ 
  154118.0-344301 &  	    	    & 13.6  & 11.83 & 10.102 &  9.373 &  9.136 & III &    \\ 
  154126.7-334732 &  	    	    & 13.69 & 12.02 & 10.144 &  9.399 &  9.154 & III &    \\ 
  154130.4-342759 &  	    	    & 13.44 & 11.81 & 10.135 &  9.279 &  9.035 & III &    \\ 
  154136.0-342532 &  	    	    & 13.58 & 11.96 & 10.336 &  9.47  &  9.197 & III &    \\ 
  154136.0-343818 &  	    	    & 13.47 & 11.83 & 10.404 &  9.747 &  9.563 & III &    \\ 
  154149.5-334517 &  	    	    & 14.13 & 12.56 & 10.863 & 10.081 &  9.775 & III &    \\ 
  154151.7-335809 &  	    	    & 13.5  & 11.52 &  9.513 &  8.746 &  8.439 & III &    \\ 
  154201.8-345449 &  	    	    & 13.48 & 11.96 & 10.511 &  9.754 &  9.536 & III &    \\ 
  154401.4-340229 &  	    	    & 13.65 & 12.09 & 10.527 &  9.658 &  9.412 & III &    \\ 
  154406.0-343238 &  	    	    & 13.66 & 12.06 & 10.602 &  9.78  &  9.577 & III &    \\ 
  154624.2-343443 &  	    	    & 13.73 & 12.0  & 10.39  &  9.611 &  9.35  & III &    \\ 
  154637.4-344307 &  	    	    & 14.69 & 12.7  & 10.713 &  9.706 &  9.317 & III &    \\  

  \noalign{\smallskip}\hline\noalign{\smallskip}
  \multicolumn{9}{c}{Lupus~3 Cloud}\\
  \noalign{\smallskip}

  160509.3-391203 &  	            & 14.0  & 12.0  & 10.571 &  9.901 &  9.667 & III &    \\
  160624.4-392158 &  	            & 13.82 & 11.77 & 10.119 &  9.385 &  9.12  & III &    \\
  160659.5-390605 &  	            & 13.98 & 11.96 & 10.496 &  9.744 &  9.525 & III &    \\
  160702.6-391203 &  	            & 14.05 & 11.86 & 10.043 &  9.072 &  8.754 & III &    \\
  160708.6-394723 &  	            & 13.5  &	    & 11.507 & 10.647 & 10.107 & II  &    \\
  160713.5-383525 &  	            & 15.25 & 13.03 & 10.719 &  9.708 &  9.372 & III &    \\
                  & Sz~93           &	    &	    & 11.089 & 10.531 & 10.352 & III &    \\
  160732.3-383508 &  	            & 13.84 & 11.74 &  9.827 &  9.034 &  8.787 & III &    \\
  160735.3-392507 &  	            & 14.17 & 12.37 & 10.88  & 10.172 &  9.993 & III &    \\
  160745.9-385245 &  	            & 13.84 & 12.18 & 10.584 &  9.905 &  9.654 & III &    \\
  160747.4-392606 &  	            & 14.03 & 12.08 & 10.333 &  9.585 &  9.373 & III &    \\
  160758.7-392109 &  	            & 14.02 & 12.08 & 10.086 &  9.282 &  8.981 & III &    \\
  160803.0-385229 &  	            & 13.88 & 12.0  &  9.514 &  8.509 &  8.098 & III &    \\
  160804.6-384558 &  	            & 13.67 & 11.72 & 10.078 &  9.413 &  9.243 & III &    \\
  160829.3-383551 &  	            & 14.65 & 12.45 & 10.285 &  9.308 &  8.838 & III &    \\
  160835.8-390348 & Par-Lup3-2      & 15.02 & 13.09 & 11.24  & 10.728 & 10.341 & III & M6 \\ 
                  & Sz~105          &	    &	    &  9.015 &  8.014 &  7.592 & III & M4 \\
  160844.3-374443 &  	            & 13.25 & 11.5  &  9.839 &  9.128 &  8.903 & III &    \\
  160846.4-393347 &  	            & 14.03 & 12.22 & 10.596 &  9.904 &  9.701 & III &    \\
  160846.6-384112 &  	            & 14.81 & 12.57 & 10.26  &  9.244 &  8.892 & III &    \\
  160855.6-392316 &  	            & 14.0  & 12.2  & 10.452 &  9.687 &  9.479 & III &    \\
  160857.3-392849 &  	            & 14.15 & 12.4  & 10.801 & 10.047 &  9.882 & III &    \\
  160903.7-385610 &  	            & 13.64 & 11.75 &  9.863 &  9.038 &  8.81  & III &    \\
  160903.8-384126 &  	            & 13.75 & 11.89 & 10.091 &  9.326 &  9.112 & III &    \\
  160917.6-392537 &  	            & 13.91 & 11.98 & 10.133 &  9.459 &  9.196 & III &    \\
  160934.1-391342 &  	            & 15.42 & 12.24 &  8.684 &  7.325 &  6.67  & III &    \\
  160937.4-391044 &  	            & 14.42 & 12.73 & 11.046 & 10.361 & 10.067 & III &    \\
  160939.5-384431 &  	            & 13.86 & 11.76 &  9.804 &  8.984 &  8.737 & III &    \\
  161013.5-384208 &  	            & 13.88 & 12.09 & 10.31  &  9.533 &  9.287 & III &    \\
  161032.6-374615 & IRAS~16072-3738 & 15.07 & 11.7  &  8.008 &  6.868 &  6.291 & III &    \\
  161033.2-383023 &  	            & 14.2  & 12.4  & 10.778 & 10.087 &  9.833 & III &    \\
  161034.5-381450 &  	            & 14.38 & 11.65 &  8.007 &  6.848 &  6.312 & III &    \\
  161036.7-380927 &  	            & 13.25 & 11.51 &  9.717 &  9.007 &  8.747 & III &    \\
  161114.8-384618 &  	            & 13.86 & 11.76 &  9.765 &  9.009 &  8.853 & III &    \\
  161118.7-385824 &  	            & 13.97 &	    &  6.79  &  5.514 &  4.842 & III &    \\
  161148.7-381758 &  	            & 15.72 & 14.99 & 13.705 & 13.036 & 12.349 & II  &    \\
  161200.1-385557 &  	            & 14.94 & 11.41 &  8.773 &  7.435 &  6.895 & III &    \\
  161200.9-383625 &  	            & 13.94 & 12.15 & 10.381 &  9.57  &  9.32  & III &    \\
  161219.6-383742 &  	            & 14.55 & 11.82 &  8.723 &  7.622 &  7.186 & III &    \\
                  & Sz~125          &	    &	    & 12.757 & 12.366 & 12.259 & III &    \\
  161251.7-384216 &  	            & 14.06 &	    &  8.566 &  7.481 &  6.988 & III &    \\
  161256.0-375643 &  	            & 14.22 & 10.98 &  7.772 &  6.637 &  6.109 & III &    \\
  161341.0-383724 &  	            & 13.22 &	    &  8.904 &  7.754 &  7.268 & III &    \\
				      						         
 \noalign{\smallskip}\hline\noalign{\smallskip} 				         
  \multicolumn{9}{c}{Lupus~4 Cloud}	\\					         
  \noalign{\smallskip}				         
  155921.8-412808 &  	            & 13.54 & 11.68 & 10.249 &  9.562 &  9.351 & III &    \\ 
  160143.3-413606 &  	            & 15.17 & 12.43 & 10.15  &  9.064 &  8.603 & III &    \\ 
  160157.0-414244 & IRAS~15585-4134 & 14.24 & 11.11 &  7.097 &  5.706 &  5.089 & III &    \\ 
\end{longtable}
\begin{flushleft}
{\bf Notes.} \\
      References: Hughes et al. (\cite{hughes94}), Mer\'{\i}n et al. (\cite{merin08}), Comer\'on (\cite{comeron08}), Comer\'on et al. (\cite{comeron09})
\end{flushleft}
}

\addtocounter{table}{1}  
\longtabL{5}{
\begin{longtable}{cc c cccc}

\caption{~~~~~~~~~~~~~~~~~~~~~~~~~~~~~Table~6: Physical parameters for Group A sources (Lupus kinematic members)}
\label{tab:grApars}\\
    \hline \hline \noalign{\smallskip}
  \multicolumn{1}{c}{SSTc2d~J} &
   \multicolumn{1}{c}{Other names} &
   \multicolumn{1}{c}{$T_{eff}$ (from spectra)} &
  \multicolumn{1}{c}{$T_{eff}$ (from SED fitting)} &
 \multicolumn{1}{c}{$L$} &
  \multicolumn{1}{c}{$M$} &
  \multicolumn{1}{c}{Age} \\
  \multicolumn{1}{l}{} &
  \multicolumn{1}{l}{} &
  \multicolumn{1}{c}{(K)} &
  \multicolumn{1}{c}{(K)} &
  \multicolumn{1}{c}{($L_{\odot}$)} &
  \multicolumn{1}{c}{($M_{\odot}$)} &
  \multicolumn{1}{c}{(Myr)} \\
      \noalign{\smallskip}\hline\noalign{\smallskip}
\endfirsthead

\caption{~~~~~~~~~~~~~~~~~~~~~~~~~~~~~Table~6: continued.}\\
   \hline\hline\noalign{\smallskip}
  \multicolumn{1}{c}{SSTc2d~J} &
   \multicolumn{1}{c}{Other names} &
   \multicolumn{1}{c}{$T_{eff}$ (from spectra)} &
  \multicolumn{1}{c}{$T_{eff}$ (from SED fitting)} &
  \multicolumn{1}{c}{$L$} &
  \multicolumn{1}{c}{$M$} &
  \multicolumn{1}{c}{Age} \\
  \multicolumn{1}{l}{} &
   \multicolumn{1}{c}{} &
  \multicolumn{1}{c}{(K)} &
  \multicolumn{1}{c}{(K)} &
  \multicolumn{1}{c}{($L_{\odot}$)} &
  \multicolumn{1}{c}{($M_{\odot}$)} &
  \multicolumn{1}{c}{(Myr)} \\
    \noalign{\smallskip}\hline\noalign{\smallskip}
\endhead
\hline
\endfoot

  \multicolumn{7}{c}{Lupus~1 Cloud}\\
  \noalign{\smallskip}
  153927.8-344617 & Sz~65, IK~Lup  	     & 3900 &	   &	   &	   &	    \\
  154009.4-342734 &                	     &      & 3400 & 0.28  & 0.4   &  1.8   \\
  154013.7-340142 &                	     &      & 3600 & 0.17  & 0.5   &  6.3   \\  			  
  154018.5-342614 &                	     &      & 3300 & 0.12  & 0.25  &  2.5   \\
  154122.0-344015 &                	     &      & 3500 & 0.12  & 0.4   &  6.4   \\
  154148.3-350145 &                	     &      & 3500 & 0.12  & 0.5   &  10.0  \\
  154517.4-341829 & Sz~69, HW~Lup  	     & 3700 & 4600 & 0.6   & 1.0   &  22.0  \\
 		  & Sz~70		     & 3100 &	   &	   &	   &	    \\
 		  & Sz~71		     & 3500 &	   &	   &	   &	    \\
 		  & Sz~72		     & 3400 &	   &	   &	   &	    \\
 		  & Sz~73		     & 3900 &	   &	   &	   &	    \\
 		  & Sz~74		     & 3600 &	   &	   &	   &	    \\
 		  & Sz~75		     & 3900 &	   &	   &	   &	    \\
 		  & Sz~76		     & 3700 &	   &	   &	   &	    \\
 		  & Sz~77		     & 3900 &	   &	   &	   &	    \\

  \noalign{\smallskip}\hline\noalign{\smallskip}
  \multicolumn{7}{c}{Lupus~3 Cloud}\\
  \noalign{\smallskip}
  		  & EX~Lup         	     & 3900 &	   &	   &	   &	    \\
  160710.1-391104 & Sz~90          	     & 3900 & 3900 & 1.1   & 1.0   &  1.4   \\
  160711.6-390348 & Sz~91          	     & 3800 & 5400 & 1.2   &	   &	    \\
  160727.1-391601 &		   	     &      & 3400 & 0.34  & 0.45  &  1.8   \\  
  160749.6-390429 & Sz~94          	     & 3300 & 3400 & 0.15  & 0.3   &  2.5   \\
  160812.6-390834 & Sz~96          	     & 3600 & 4000 & 0.93  & 1.1   &  2.9   \\
  160817.4-390105 &		   	     &           & 3100 & 0.15  & 0.2   &  1.3   \\
  160821.8-390422 & Sz~97, Th~24   	     & 3400 & 3200 & 0.19  & 0.25  &  1.3   \\
  160822.5-390446 & Sz~98, HK~Lup  	     & 3900 &	   &	   &	   &	    \\
  160822.8-390058 &		   	     &      & 3400 & 0.25  & 0.45  &  2.8   \\
  160825.2-384055 &		   	     &      & 3300 & 0.16  & 0.25  &  1.8   \\
  160825.8-390601 & Sz~100, Th~26  	     & 3100 & 2900 & 0.24  &	   & $<1.0$ \\
  160827.8-390040 &	           	     &      & 3300 & 0.3   & 0.3   &  1.0   \\
  160828.4-390532 & Sz~101, Th~27  	     & 3300 & 2900 & 0.39  &	   & $<1.0$ \\
  160829.7-390311 & Sz~102, Krautter's star  & 5000 & 4400 & 0.02  &	   &	    \\
  160830.3-390611 & Sz~103, Th~29            & 3300 & 3000 & 0.17  & 0.15  &  1.3   \\
  160839.8-390625 & Sz~106                   & 3500 & 4600 & 0.59  & 1.05  &  22.4  \\
  160839.8-392922 &		             &      & 3200 & 0.11  & 0.25  &  2.8   \\
  160841.8-390137 & Sz~107                   & 3000 & 3000 & 0.18  & 0.15  &  1.3   \\
  160851.6-390318 & Sz~110, Th~32            & 3200 & 4400 & 0.5   & 1.05  &  20.0  \\
  160853.2-391440 & 2MASS~J16085324-3914401  &      & 3900 & 0.44  & 0.95  &  6.4   \\
  160854.7-393744 & Sz~111, Th~33            & 3600 &	   &	   &	   &	    \\
  160855.5-390234 & Sz~112                   & 3000 & 3000 & 0.24  &	   & $<1.0$ \\
  160901.8-390513 & Sz~114, V908~Sco         & 3000 & 2900 & 0.42  &	   & $<1.0$ \\
  160906.2-390852 & Sz~115                   & 3300 & 3100 & 0.17  & 0.175 &  1.0   \\
  160915.7-385139 &		             &      & 3400 & 0.083 & 0.35  &  8.0   \\
  160942.6-391941 & Sz~116, Th~36            & 3600 & 5400 & 1.5   &	   &	    \\
  160944.3-391330 & Sz~117, Th~37            & 3600 & 3700 & 0.47  & 0.75  &  3.2   \\
  160948.6-391117 & Sz~118                   & 4100 & 4800 & 4.3   & 1.4   &  11.3  \\
  160957.1-385948 & Sz~119                   & 3300 & 3100 & 0.41  &	   & $<1.0$ \\
  161012.2-392118 & Sz~121, Th~40            & 3400 & 3300 & 0.67  & 0.45  &  1.0   \\
  161034.3-381031 &		             &      & 3400 & 0.42  & 0.45  &  1.1   \\
  161041.9-382304 &		             &      & 3300 & 0.24  & 0.35  &  1.8   \\
  161051.6-385314 & Sz~123, Th~42            & 3400 & 3900 & 0.4   & 0.9   &  6.4   \\
  161138.1-384135 &		             &      & 3400 & 0.25  & 0.45  &  2.8   \\
  161153.4-390216 & Sz~124, Th~43            & 3900 &	   &	   &	   &	    \\
  161207.6-381324 &		             &      & 3400 & 0.26  & 0.4   &  2.0   \\

 \noalign{\smallskip}\hline\noalign{\smallskip}
 \multicolumn{7}{c}{Lupus~4 Cloud}\\
 \noalign{\smallskip}
		  & Sz~128                   & 3600 &	   &	   &	   &	    \\
		  & Sz~129                   & 3900 &	   &	   &	   &	    \\
  160031.1-414337 & Sz~130                   & 3600 & 3900 & 0.26  & 0.9   &  17.9  \\  
  160049.4-413004 & Sz~131                   & 3500 & 4000 & 0.15  & 0.75  &  39.8  \\
  160329.4-414003 & Sz~133                   & 4700 & 4400 & 0.36  & 0.95  &  25.2  \\
\end{longtable}
}

\end{document}